\documentclass{JHEP3}

\usepackage{enumerate}
\usepackage{amsmath}
\usepackage{amssymb}
\usepackage{amsfonts}
\usepackage{latexsym}
\usepackage{epsfig}
\usepackage{array}
\usepackage{multirow}
\usepackage[vcentermath]{youngtab}

\newcommand\field[1]{{\ensuremath{\mathbb{{#1}}}}}

\newcommand{\RR}{\field{R}}

\newcommand{\ZZ}{\field{Z}}

\newcommand{\sT}{\mathcal{T}}
\newcommand{\GG}{\mathcal{G}}
\newcommand{\MM}{\mathcal{M}}
\newcommand{\NN}{\mathcal{N}}
\newcommand{\PP}{\mathcal{P}}

\newcommand{\tF}{\tilde{F}}
\newcommand{\p}{\partial}

\newcommand{\OO}{\mathcal{O}}
\newcommand{\HH}{\mathcal{H}}
\newcommand{\ta}{\tilde{\alpha}}

\def\ov{\over}

\def\lam{{\lambda}}

\def\det{{\rm det}}
\def\tr{{\rm tr}}
\def\Tr{{\rm Tr}}
\def\NN{{\cal N}}

\def\eq#1{(\ref{#1})}
\def\modular#1{{\quad (\text{mod }#1)}}

\def\Om{{\Omega}}
\def \th{{\theta}}

\def \lam {\lambda}
\def \om {\omega}

\def\LL{{\cal L}}

\def\n{{V_A}}

\def\LL{{\cal L}}
\def\NN{{\cal N}}

\newcommand{\be}{\begin{equation}}
\newcommand{\ee}{\end{equation}}
\newcommand{\bea}{\begin{eqnarray}}
\newcommand{\eea}{\end{eqnarray}}

\newcommand{\bln}{\begin{align}}
\newcommand{\eln}{\end{align}}
\newcommand{\bst}{\begin{split}}
\newcommand{\est}{\end{split}}
\newcommand{\bi}{\begin{itemize}}
\newcommand{\ei}{\end{itemize}}
\newcommand{\ben}{\begin{enumerate}}
\newcommand{\een}{\end{enumerate}}

\title{Constraints on 6D Supergravity Theories \\
with Abelian Gauge Symmetry}

\author{Daniel S. Park and Washington Taylor\\
Center for Theoretical Physics\\
Department of Physics\\
Massachusetts Institute of Technology\\
Cambridge, MA 02139, USA\\
\\
\\
{\tt dspark81} {\rm at} {\tt mit.edu},
{\tt wati} {\rm at} {\tt mit.edu}
}

\preprint{MIT-CTP-4228}

\abstract{We study six-dimensional $\NN=(1,0)$
supergravity theories with abelian, as well as non-abelian, gauge group factors.
We show that for theories with fewer than nine tensor multiplets,
the number of possible combinations of
gauge groups---including abelian factors---and non-abelian matter representations is finite.
We also identify infinite families of theories with distinct $U(1)$
charges that cannot be ruled out using known quantum consistency conditions,
though only a finite subset of these can arise from known string constructions.}

\begin{document}

\section{Introduction}

Using quantum consistency conditions to identify interesting theories
among a general class of theories has proven to be a fruitful practice
historically.  This is epitomized by the sequence of developments that
initiated with the classic works by Alvarez-Gaum\'e and Witten
\cite{ganomalies} and by Green and Schwarz \cite{GS} on anomalies, and
that culminated in the discovery of the heterotic string by Gross,
Harvey, Martinec and Rohm \cite{heterotic}.  Gauge, gravitational and
mixed anomalies \cite{ganomalies,anomalies} constrain the gauge
symmetry of ten-dimensional supergravity theories.  The anomalies
cannot cancel unless the gauge group is one of the groups $SO(32)$,
$E_8 \times E_8$, $E_8 \times U(1)^{248}$ or $U(1)^{496}$\cite{GS,
GSW}.  The $SO(32)$ type I string theory was known at the time that
these constraints were discovered, but this result motivated the
discovery of the $E_8 \times E_8$ and $Spin(32)/ \ZZ_2$ heterotic
string theories.  Recent results show that the $E_8 \times U(1)^{248}$
and $U(1)^{496}$ theories cannot be made simultaneously anomaly-free
and consistent with supersymmetry \cite{Adams:2010zy}.  
Together,
these results show that the spectra of all anomaly-free gravity
theories with minimal supersymmetry in 10D are realized in string
theory.  

Gravitational anomaly constraints were also used
to identify the unique massless spectrum that a six-dimensional
supergravity theory with $\NN=(2,0)$ symmetry can have \cite{6d20}.
This spectrum precisely agrees with the massless spectrum of
type IIB string theory compactified on a K3 manifold.
These results suggest that quantum
consistency conditions---especially anomaly constraints---are
an effective tool for obtaining clues about the microscopic theory of
gravity from the structure of the macroscopic theory.

An interesting class of theories to approach from this perspective
is the set of six-dimensional $\NN=(1,0)$ supergravity theories.
These theories have strong anomaly constraints, but at the same time admit diverse
consistent string vacua with a wide range of gauge groups
and matter representations.
A sampling of the substantial
literature on six-dimensional $\NN=(1,0)$ supergravity theories and
their
string realizations is given in references
\cite{Nishino-Sezgin}-\cite{Morrison-Taylor};
reviews of this work and further references appear in
\cite{6Dreview}, \cite{TaylorTASI}.
Constraints on the set of nonabelian gauge groups
and matter that can appear in such theories were
analyzed in \cite{Avramis:2005hc,Kumar:2009us,
Kumar:2009ae,Kumar:2009ac,Kumar:2010ru,Taylor:2010wm,
Kumar:2010am}.
We are interested here in obtaining constraints on these theories
in the case when there are abelian vector fields
in the massless spectrum.

Many six-dimensional $(1,0)$ supergravity theories with abelian
vector fields have been identified as arising from string theory
vacuum constructions.
Erler has analyzed the $U(1)$ sector of heterotic orbifold constructions 
in \cite{Erler:1993zy} while Honecker and Trapletti have extended the
analysis to more general heterotic backgrounds \cite{Honecker}.
Abelian gauge symmetry of six-dimensional F-theory backgrounds
has also
been studied by various authors \cite{Aldazabal,Berglund,Aspinwall}.
Determining the abelian gauge symmetry
of a given string model is often subtle because
vector bosons in the spectrum that are naively
massless can be lifted at the linear level by
coupling to St\"uckelberg fields.
This is a generic phenomenon
in many types of string models including heterotic,
orbifold, intersecting brane,
fractional brane and F-theory models\footnote{We have given
an incomplete list of references that address this
in the bibliography \cite{GSW6,Strominger,Erler:1993zy,
MV,Berkooz:1996iz,Aldazabal,Berglund,Aspinwall,Honecker,
Blumenhagen,Buican:2006sn,Grimm,Hayashi,Joeetal}.}.
Therefore, a detailed analysis of all the vector-matter couplings
in play is necessary in order to ensure that a vector boson is
indeed in the massless spectrum in a given string model.

The anomaly cancellation structure for specific six-dimensional
$(1,0)$ string vacua with abelian gauge symmetry has been worked out
in the literature
\cite{Erler:1993zy,Aldazabal,Berglund,Aspinwall,Honecker}.  General
analyses on the full landscape of six-dimensional $(1,0)$ supergravity
theories, however, have not fully incorporated the structure of
abelian gauge symmetries.  In this paper we treat the abelian anomaly
cancellation conditions carefully and show how they place bounds on
the massless spectrum of general six-dimensional supergravity theories
with abelian factors in the gauge group.



It was shown in \cite{Kumar:2009ae,Kumar:2010ru} that the number of non-abelian theories
with $T<9$ tensor multiplets is
finite.
In light of this result it is natural to ask whether
this bound continues to hold when we allow abelian gauge group factors.
It is useful to divide this question into two parts.
That is, we ask;
\begin{enumerate}[I.]
\item Whether the number of different gauge/matter structures
is finite when we ignore the charges of the matter
under the $U(1)$'s.
\item Whether given the gauge/matter
structure, the number of distinct combinations of $U(1)$ charges
each matter multiplet can have is finite.
\end{enumerate}
In this paper we show that the answer to the first question is
``yes" when $T<9$.
As is the case with non-abelian theories,
when $T \geq 9$ we can generate an infinite class of theories
in which the bounds that hold for $T<9$ theories are violated.
An example of such an infinite class is given in section \ref{ss:Tgeq9}.

In addressing the second question,
it is important to note that theories with multiple $U(1)$
gauge symmetries (say $U(1)^n$) are defined up to
arbitrary linear redefinitions of the gauge
symmetry. If we assume that all the $U(1)$'s
are compact and normalize the unit charge to be $1$
for each $U(1)$ factor, the theories are defined up to $SL(n,\ZZ)$.

From this fact, we may deduce that there are an infinite number of
distinct $U(1)$ charge assignments possible for certain non-anomalous
gauge/matter structures.  This is because there are many known
examples of theories with two $U(1)$ factors and at least one
uncharged scalar, so that the non-anomalous gauge group can be written
in the form $U(1)^2 \times \GG_0$.  Since any linear combination of
the two $U(1)$'s is a non-anomalous $U(1)$ gauge symmetry, it is
possible to construct an infinite class of apparently consistent 6D
supergravity theories with gauge group $U(1) \times \GG_0$ by simply
removing the other $U(1)$ along with a neutral scalar from the
spectrum.


Hence we see that the answer to the
second question is negative. We may now ask, however,

\begin{enumerate}[I.]
\setcounter{enumi}{2}
\item Whether all infinite families of $U(1)$'s could be
generated in the trivial manner presented above.
\item Whether additional quantum consistency conditions
that are unknown to us at the present
could be employed to constrain the set of
$U(1)$ charges in a given theory.
\end{enumerate}
Regarding question III, we find that there are non-trivially generated infinite families
of $U(1)$ charge solutions.
We have not addressed the last question here, though some speculation
in this regard is included at the end of the conclusions in Section
\ref{s:conclusion}.


This paper is organized as follows:
In section \ref{s:u1anom} we review the massless spectrum
and anomaly structure of six-dimensional $(1,0)$ theories.
In section \ref{s:fin} we address the first question.
In section \ref{s:inf} we address the second and third questions.
In particular, we present examples of
infinite classes of $T=1$ theories with $U(1)$'s
that are trivially/nontrivially generated.
We also discuss  subtleties arising in the case $T=0$, where there are
no tensor multiplets.
We conclude our discussion in section \ref{s:conclusion}.

The results in this paper are based on consistency conditions on
low-energy supergravity theories, and do not depend upon a specific UV
completion such as string theory.  In some cases, however, examples
are drawn from string theory and F-theory to illuminate the structure
of the set of allowed models as determined from macroscopic
considerations.

\section{6D $(1,0)$ Theories and Anomaly Cancellation} \label{s:u1anom}

In this section we review six-dimensional theories
with $\NN=(1,0)$ supersymmetry
and anomaly cancellation in these theories.
In section \ref{ss:6D} we present an overview
of the field content of these theories.
We compute the anomaly polynomial in section \ref{ss:poly}
and review  anomaly cancellation and factorization
in section \ref{s:comments}. We give explicit formulae for the anomaly
factorization
condition 
in the presence of $U(1)$'s in section \ref{ss:factcond}
and discuss some salient features of these equations.

In sections \ref{ss:teq1} and \ref{ss:teq0}
we summarize the aforementioned aspects of anomaly cancellation
specializing to the cases of $T=1$ and $T=0$ respectively.
In section \ref{ss:Stuckelberg} we discuss aspects of
the generalized Green-Schwarz mechanism that come into play
when the theory has abelian gauge symmetry, and also
explain why this issue can be safely ignored when
discussing the massless spectrum.

\subsection{The Massless Spectrum} \label{ss:6D}

The massless spectrum of the models we consider
can contain four different multiplets of the supersymmetry algebra:
the gravity and tensor multiplet,
vector multiplet, and hypermultiplet. The contents of
these multiplets are summarized in Table \ref{t:mult}.

We consider theories with one gravity multiplet.
There can in general be multiple tensor multiplets;
we denote the number of
tensor multiplets by $T$. When $T=1$ it is possible to
write a Lagrangian for the theory;
the self-dual and anti-self-dual tensors
can combine into a single antisymmetric tensor.
Theories with $T$ tensor multiplets have a moduli space with
$SO(1,T)$ symmetry; the $T$ scalars in each multiplet
combine into a $SO(1,T)$ vector $j$ that can be taken to have
unit norm. We consider theories with arbitrary gauge group
and matter content.

Note that a theory with a general number of tensor multiplets
can still be defined despite the lack of a covariant Lagrangian.
The partition function can be defined by coupling the three-form field
strength to a 3-form gauge potential as in \cite{kinetic}.
Classical equations of motion can be formulated as in
 \cite{Romans,Sagnotti:1992qw}.
Supersymmetry and anomaly cancellation may be discussed
at the operator level of a theory obtained by quantizing the
classical theory defined by these equations.


We write the gauge group for a given theory as\footnote{The 
gauge group generally can have a quotient by a discrete subgroup,
but this does not affect the gauge algebra, which underlies
the anomaly structure analyzed in this paper.}
\be
\GG = \prod_{\kappa=1}^\nu \GG_\kappa \times \prod_{i=1}^{\n} U(1)_i \,.
\ee
Lowercase greek letters $\kappa,\lambda,\cdots$ are
used to denote the simple non-abelian gauge group factors;
lowercase roman letters $i,j,k,\cdots$ are used to denote $U(1)$
factors. $\nu$ and $V_A$  denote the numbers of
nonabelian and abelian gauge group factors of the theory.

We denote by $N$ the number of irreducible representations
of the non-abelian gauge group under which the matter hypermultiplets
transform (including trivial representations);
we use uppercase roman letters to index these representations.
We say that hypermultiplets transform in
the representation $R^I_\kappa$ under $\GG_\kappa$
and have $U(1)_i$ charge $q_{I,i}$.

\begin{table}[t!]
\center
 \begin{tabular}{ | c | c |}
 \hline
 Multiplet & Field Content\\ \hline
 Gravity & $(g_{\mu \nu}, \psi^+_\mu, B^+_{\mu \nu}) $  \\ \hline
 Tensor & $(\phi, \chi^-, B^-_{\mu \nu})$   \\ \hline
 Vector & $(A_{\mu}, \lam^+)$   \\ \hline
 Hyper & $(4\varphi, \psi^-)$  \\ \hline
  \end{tabular}
 \caption{Six-dimensional (1,0) supersymmetry multiplets. The signs on
  the fermions indicate the chirality. The signs on antisymmetric tensors
  indicate self-duality/anti-self-duality.}
\label{t:mult}
\end{table}

We characterize theories by their massless spectrum.
There is a slight subtlety we must consider when dealing with
$U(1)$ gauge symmetries. It is possible to break $U(1)$ at the
linearized level by certain hypermultiplets, called
``linear hypermultiplets" in the literature \cite{Howe:1983fr}.
We will refer to these multiplets simply as ``linear multiplets"
throughout this paper.
When a linear multiplet couples to a vector multiplet the two merge into
a long (or non-BPS) multiplet and are lifted from the massless
spectrum. Once lifted from the massless spectrum,
these long multiplets can be safely ignored.
This issue is discussed in more detail in section \ref{ss:Stuckelberg}.

\subsection{The Anomaly Polynomial for Theories with $U(1)$'s} \label{ss:poly}

In six-dimensional chiral theories there can be gravitational, gauge and
mixed anomalies \cite{anomalies}. The sign with which each
chiral field contributes to the anomaly is determined
by their chirality.

The 6D anomaly can be described by the method of descent from
an 8D anomaly polynomial. The anomaly polynomial
is obtained by adding up the contributions of all the
chiral fields present in the theory  \cite{ganomalies}. For the $T=1$ case this is given in
 \cite{Erler:1993zy,Honecker}.
In general we obtain
\begin{align}
\label{raw}
\begin{split}
I_8=
&-{1 \ov 5760} (H-V+29T-273) [\tr R^4+{5 \ov 4} (\tr R^2)^2] \\
&-{1\ov 128} (9-T) (\tr R^2)^2 \\
&-{1 \ov 96} \tr R^2 [\sum_\kappa \Tr F_\kappa^2
 -\sum_{I,\kappa} \MM^\kappa_I \tr_{R^I_\kappa}F_\kappa^2] \\
&+{1\ov 24} [\sum_\kappa \Tr F_\kappa^4
 -\sum_{I,\kappa} \MM^\kappa_I \tr_{R^I_\kappa}F_\kappa^4
 -6\sum_{I,\kappa,\lambda} \MM^{\kappa\lambda}_I
 (\tr_{R^I_\kappa}F_\kappa^2)(\tr_{R^I_\lambda}F_\lambda^2)] \\
 &+{1 \ov 96} \tr R^2 \sum_{I,i,j} \MM_I q_{I,i}q_{I,j} F_i F_j \\
&-{1 \ov 6} \sum_{I,\kappa,i} \MM^\kappa_I q_{I,i}
 (\tr_{R^I_\kappa}F_\kappa^3) F_i 
-{1 \ov 4} \sum_{I,\kappa,i, j} \MM^\kappa_I q_{I,i} q_{I,j}
 (\tr_{R^I_\kappa}F_\kappa^2) F_i F_j \\
&-{1 \ov 24}\sum_{I,i,j,k,l} \MM_I q_{I,i} q_{I,j} q_{I,k}q_{I,l} F_i F_j F_k F_l \,.
\end{split}
\end{align}
$\MM_I$ is the size of the representation $I$
that is given by
\be
\MM_I = \prod_\kappa d_{R_\kappa^I} \,,
\ee
where $d_{R_\kappa}$ is the dimension of the representation
$R_\kappa$ of $\GG_\kappa$.
Similarly, $\MM^\kappa_I$ ($\MM^{\kappa\lambda}_I$)
is the number of $\GG_\kappa$ ($\GG_\kappa \times \GG_\lambda$)
representations in $I$, which is given by
\be
\MM^\kappa_I = \prod_{\mu \neq \kappa} d_{R_\mu^I} \qquad
(\MM^{\kappa\lambda}_I = \prod_{\mu \neq \kappa,\lambda} d_{R_\mu^I})
\ee
respectively. $V$ and $H$ are the number of
massless vector multiplets and hypermultiplets in the theory.
They are given by
\be
V \equiv V_{NA}+V_A \equiv \sum_\kappa d_{\text{Adj}_\kappa}+V_A,
\qquad H \equiv \sum_I \MM_I
\ee
where $d_{\text{Adj}_\kappa}$ is the dimension of the adjoint
representation of gauge group $\GG_\kappa$.
$V_{NA}$ is the number of non-abelian vector multiplets in the theory.
The integer $N$, which is the number of
irreducible representations of the non-abelian gauge group, plays an
important role in bounding the number of $U(1)$'s.
We use `$\tr$' to denote the trace in the fundamental representation,
and `$\Tr$' to denote the trace in the adjoint. Multiplication of forms should be interpreted
as wedge products throughout this paper unless stated otherwise.

\subsection{Anomaly Cancellation and Factorization} \label{s:comments}

The Green-Schwarz mechanism \cite{GS} can be generalized to
theories with more than one tensor multiplet when the anomaly polynomial
is factorizes in the following form \cite{Sagnotti:1992qw,Sadov}:
\be
\label{genfactorization0}
I_8=-{1 \ov 32} \Om_{\alpha \beta} X_4^\alpha X_4^\beta \,,
\ee
where $\Om$ is a symmetric bilinear form (or metric) in
$SO(1,T)$ and $X_4$ is a four form that is an $SO(1,T)$ vector.
$X_4$ can be written as
\be
X_4^\alpha = {1 \ov 2} a^\alpha \tr R^2
+ \sum_\kappa ({2b_\kappa^\alpha \ov \lambda_\kappa})
\tr F_\kappa^2 + \sum_{ij} 2 b^\alpha_{ij} F_i F_j \,,
\ee
where we define $b_{ij}$ to be symmetric in $i,j$. The $a$
and $b$'s are $SO(1,T)$ vectors and $\alpha$ are $SO(1,T)$ indices.
Note that the anomaly coefficients for the $U(1)$'s 
can be written in this way due to the fact that the field
strength is gauge invariant on its own \cite{Riccioni}.
The $\lambda_\kappa$'s are normalization factors that are fixed
by demanding that the smallest topological charge of an embedded
SU(2) instanton is 1. These factors, which are equal to
the Dynkin indices of the fundamental representation of each gauge group,
are listed in table \ref{t:lambda} for all the simple groups.
The $b_\kappa$'s form an integral
$SO(1,T)$ lattice when we include these normalization
factors \cite{Kumar:2010ru}.

\begin{table}[b!]
\center
 \begin{tabular}{ | c | c | c | c | c | c | c | c | c | c |}
 \hline
 & $A_n$ & $B_n$ & $C_n$ & $D_n$ & $E_6$ &$E_7$ & $E_8$ & $F_4$ & $G_2$ \\ \hline
 $\lambda$ & 1 &2& 1& 2& 6 &12& 60 &6 &2 \\ \hline
  \end{tabular}
 \caption{Normalization factors for the simple groups.}
\label{t:lambda}
\end{table}

The gauge-invariant three-form field strengths are given by
\be
H^\alpha=dB^\alpha+{1 \ov 2} a^\alpha \om_{3L}
+2\sum_\kappa {b^\alpha_\kappa \ov \lambda_\kappa} \om^\kappa_{3Y}
+2\sum_{ij} {b^\alpha_{ij}} \om^{ij}_{3Y},
\ee
where $\om_{3L}$ and
$\om_{3Y}$ are Chern-Simons 3-forms of the spin connection
and gauge fields respectively. If the factorization condition
(\ref{genfactorization0}) is satisfied, anomaly cancellation can
be achieved by adding the local counterterm
\be
\delta \LL_{GS} \propto -\Om_{\alpha\beta} B^\alpha \wedge X_4^\beta \,.
\ee

Meanwhile, supersymmetry determines the kinetic term for the gauge
fields to be (up to an overall factor) \cite{Sagnotti:1992qw,Riccioni}
\be
-\sum_\kappa ({j \cdot b_\kappa \ov \lambda_\kappa}) \tr (F_\kappa \wedge * F_\kappa)
-\sum_{ij} (j \cdot b_{ij}) (F_i \wedge * F_j) \,,
\ee
where $j$ is the unit $SO(1,T)$ vector that parametrizes
the $T$ scalars in the tensor multiplets.
The inner product of $j$ and the $b$ vectors are defined
with respect to the metric $\Om$.  There must be a value of
$j$ such that all the gauge fields have positive definite kinetic terms.
This means that there should be some value of $j$ such
that all $j \cdot b_\kappa$
are positive and such that $j \cdot b_{ij}$ is a positive definite matrix with
respect to $i,j$.

If we did not have any $U(1)$'s, (\ref{genfactorization0}) would be
the only way in which the anomaly can be cancelled. When we have
abelian vector multiplets, however, a generalized version of the
Green-Schwarz mechanism is available \cite{Berkooz:1996iz}.  In this
case, it is possible to cancel terms in the 8-form anomaly polynomial
that are proportional to \be F \wedge X_6 \,, \ee where $X_6$ is a six
form, by a counter-term in the action of the form \be -C \wedge X_6
\,, \ee where $C$ is a St\"uckelberg 0-form that belongs to a linear
multiplet.  The coupling of $C$ to the vector boson $V$ is given by
\be {1 \ov 2}(\p_\mu C- V_\mu )^2 \,, \ee which is what we mean by $C$
being a St\"uckelberg 0-form.  The anomalous gauge boson $V$ recieves
a mass, hence rendering the $U(1)$ broken; the abelian vector
multiplet is lifted from the massless spectrum by coupling to the
linear multiplet by the St\"uckelberg mechanism. When all the
anomalous $U(1)$'s are lifted and we look at the pure massless
spectrum of the theory, all the gravitational anomalies and
gauge/mixed anomalies induced by the massless fields are cancelled
completely by two forms through the conventional Green-Schwarz
mechanism.  The lesson is that when we are discussing the massless
spectrum, this generalized version of the Green-Schwarz mechanism does
not come into play and can be safely ignored.\footnote{The situation
is quite the opposite when we are taking the top-down approach, for
example when we are constructing theories from string
compactifications.  Since we are working downward from the high-energy
end, it is then important to figure out which $U(1)$ vector bosons
that naively seem to be massless are lifted by this mechanism.}  We
elaborate further
on this issue in section \ref{ss:Stuckelberg}.

\subsection{The Factorization Equations} \label{ss:factcond}


We are now ready to write down the factorization equations
in the presence of $U(1)$'s.
The factorization equations come from demanding that the anomaly
polynomial \eq{raw} factorize in the form
\eq{genfactorization0}.
Comparing
the terms with no abelian field strength factors gives the conditions
\begin{align}
R^4~&: &273&=H-V+29T \label{grav}\\
(R^2)^2~&: &a\cdot a&= 9-T \label{nab}\\
F^2R^2~&: &a\cdot b_\kappa&=
{1 \ov 6} \lambda_\kappa (A_{\text{Adj}_\kappa}-\sum_I \MM_I^\kappa A^I_\kappa)\\
F^4~&: &0&=
B_{\text{Adj}_\kappa}-\sum_I \MM_I^\kappa B^I_\kappa \\
(F^2)^2~&: &b_\kappa \cdot b_\kappa&=
{1 \ov 3} \lambda_\kappa^2 (\sum_I \MM_I^\kappa C^I_\kappa-C_{\text{Adj}_\kappa})\\
F_\kappa^2 F_\mu^2~&: &b_\kappa \cdot b_\mu&=
\lambda_\kappa \lambda_\mu \sum_I \MM^{\kappa\mu}_I A^I_\kappa A^I_\mu \label{nae}
\end{align}
The inner products on the left-hand-side of the equations
are taken with respect to the $SO(1,T)$ metric $\Om$.
For each representation $R$ of
a given group, the group theory coefficients $A_R,B_R,C_R$
are defined by
\be
\tr_R F^2 = A_R \tr F^2,\qquad
\tr_R F^4 = B_R \tr F^4 + C_R (\tr F^2)^2 \,,
\ee
where $\tr$ denotes the trace with respect to the fundamental
representation. For each hypermultiplet $I$ we use the
shorthand notation
\be
A^I_\kappa=A_{R^I_\kappa},\quad
B^I_\kappa=B_{R^I_\kappa},\quad
C^I_\kappa=C_{R^I_\kappa} \,.
\ee
We refer to the first anomaly equation \eq{grav} as the
gravitational anomaly constraint.

The $U(1)$ anomaly equations obtained by comparing terms with
abelian field strength factors are given by
\[
\fbox{
\addtolength{\linewidth}{-2\fboxsep}%
\addtolength{\linewidth}{-2\fboxrule}%
\begin{minipage}{\linewidth}
\begin{align}
F_iF_jR^2~&: &
a \cdot b_{ij} &= -{1 \ov 6} \sum_I \MM_I q_{I,i} q_{I,j} \label{one} \\
F_i^3 F_\kappa~&: &
0 &= \sum_I \MM_I^\kappa E^I_\kappa q_{I,i} \\
F_iF_jF_\kappa^2~&: &
({b_\kappa \ov \lambda_\kappa}) \cdot b_{ij} &=
\sum_I \MM_I^{\kappa} A^I_\kappa q_{I,i}q_{I,j} \label{three}\\
F_iF_jF_kF_l~&: &
b_{ij} \cdot b_{kl} + b_{ik} \cdot b_{jl} + b_{il} \cdot b_{jk}
&= \sum_I \MM_I q_{I,i} q_{I,j} q_{I,k} q_{I,l} \label{four}
\end{align}
\end{minipage}\nonumber
}
\]
for all $i,j,k,l$.
The group theory coefficient $E$ is defined to be
\be
\tr_R F^3 = E_R \tr F^3
\ee
and $E^I_\kappa =E_{R^I_\kappa}$. 

It is useful to summarize these anomaly constraints 
by the following polynomial identities;
\[
\fbox{
\addtolength{\linewidth}{-2\fboxsep}%
\addtolength{\linewidth}{-2\fboxrule}%
\begin{minipage}{\linewidth}
\begin{align}
a\cdot P(x_i)&= -{1 \ov 6} \sum_I \MM_I f_I(x_i)^2 \label{1st}\\
0&=\sum_I \MM_I^\kappa E^I_\kappa f_I (x_i) \label{2nd}\\
b_\kappa \cdot P(x_i)&=
\lambda_\kappa \sum_I \MM_I^{\kappa} A^I_\kappa f_I(x_i)^2 \label{3rd}\\
P(x_i) \cdot P(x_i)&= {1 \ov 3} \sum_I \MM_I f_I(x_i)^4 \label{4th}
\end{align}
\end{minipage}\nonumber
}
\]
Here we have defined the $SO(1,T)$ scalar and vector polynomials
\begin{align}
f_I (x_i) &\equiv \sum_i q_{I,i}x_i \\
P^\alpha (x_i) &\equiv \sum_i b_{ij}^\alpha x_i x_j \,. \label{eq:pa}
\end{align}
The reason that $U(1)$ factorization conditions can be written as polynomial
identities is because the field strengths of the $U(1)$'s behave like numbers
rather than matrices in the anomaly polynomial.
We note that the $x_i$ are auxiliary variables
and do not have any physical significance.


A theory with charges $q_{I,i}$ assigned to the hypermultiplets
is only consistent if there exist $b_{ij}$ satisfying these equations
that give a positive-definite kinetic matrix
$j \cdot b_{ij}$ for the $U(1)$ gauge fields.
It is useful to define the charge vector with respect to $U(1)_i$
whose components are the charges of the $N$ nonabelian representations:
\be
\vec{q}_i \equiv (q_{1,i}, q_{2,i}, \cdots, q_{N,i})
\ee

There is a $GL(\n,\field{R})$ symmetry of the
$U(1)$ anomaly equations that originates from the fact that there
is a freedom of redefining $U(1)$'s.
If there are multiple $U(1)$'s one could take
some new linear combination of them to define
a new set of non-anomalous $U(1)$'s.
The equations are invariant under
\be
\begin{pmatrix} \vec{q_1} \\ \vec{q_2} \\ \vdots \\ \vec{q_\n} \end{pmatrix} \rightarrow
M \begin{pmatrix} \vec{q_1} \\ \vec{q_2} \\ \vdots \\ \vec{q_\n} \end{pmatrix},
\quad
(b^\alpha)_{ij} \rightarrow  \left(M^{t}(b^\alpha) M \right)_{ij},
\label{eq:invariance}
\ee
for $M \in GL(\n,\field{R})$.
We have denoted $(b^\alpha)$ to be the matrix whose
$(i,j)$ element is $b^\alpha_{ij}$.
When we are discussing properly quantized charges of
compact $U(1)$'s the linear redefinitions of the $U(1)$'s
must be given by elements of $SL(\n,\field{Z}) \subset GL(\n,\field{R})$.
Here, however, we merely use the fact that the anomaly equations
are invariant under $GL(\n,\field{R})$ as a tool for obtaining bounds
on the number of $U(1)$'s we can add to a given theory.
Therefore, we do not need to be concerned with the issue of integrality of charges.

The factorization equations, combined with the positive-definite
condition on $b_{ij}$, impose stronger constraints on the theory 
when $T<9$. This is because $a$ is
timelike when $T<9$:
\be
a\cdot a = 9-T >0
\ee
When $a$ is timelike,
\be
a\cdot y=0, ~~ y \cdot y \geq 0 \qquad \Rightarrow \qquad y=0
\ee
for any arbitrary $SO(1,T)$ vector $y$. This fact is
used in \cite{Kumar:2010ru} to bound the number of theories with
nonabelian gauge groups, and is also crucial in bounding
the space of theories with abelian factors.
In particular, this fact implies that
the charge vectors $\vec{q}_i$ must be linearly independent
in order to get a positive definite kinetic term for the $U(1)$'s when $T<9$.
If they are not, there exists non-zero $(x_i)$ such that $f_I(x_i)=0$
for all $I$ since
\be
\vec{f}(x_i) \equiv (f_1(x_i),\cdots,f_N(x_i)) = \sum_i x_i \vec{q}_i \,.
\ee
For such $x_i$, we see that
\be
a\cdot P(x_i) =0, \qquad P(x_i) \cdot P(x_i)=0 \,.
\ee
This implies that $P(x_i)=0$, which in turn implies that $j \cdot P(x_i)=0$,
{\it i.e.,}
\be
\sum_{ij} (j \cdot b)_{ij} x_i x_j = 0 \,.
\ee
This would mean that the kinetic term is not positive-definite.
Hence we have proven that in order for the kinetic term to be
positive-definite, $\vec{q}_i$ must be linearly independent when $T < 9$.
This in particular means that we cannot have a massless $U(1)$
vector under which nothing is charged, {\it i.e.,} that when $T<9$,
the trivial solutions to the $U(1)$ factorization equations
where all the charges are set to $q_{I,i}=0$ are not acceptable.
The analogous connection in 10D between $U(1)$ charges
and the $BF^2$ term, which is related by supersymmetry
to the gauge kinetic term \cite{Sagnotti:1992qw},
also played a key role in the analysis in
 \cite{Adams:2010zy} showing that the ten-dimensional supergravity
theories with gauge group $U(1)^{496}$ and $E_8 \times U(1)^{248}$
are inconsistent.

The fact that $\vec{q}_i$ are all linearly independent for $T <9$
also implies that
\be
\label{positivity}
P(x_i)\cdot P(x_i) = \sum_I \MM_I f_I(x_i)^4 >0
\ee
for all non-zero $x_i$ as $f_I(x_i)$ cannot be made simultaneously
zero for all $I$.
We make use of \eq{positivity} in bounding the set of abelian theories
in section \ref{s:fin}.

\subsection{$T=1$} \label{ss:teq1}

In this section we discuss anomaly cancellation and the factorization
equations in the special case of one tensor multiplet.  As discussed
earlier, these theories have a Lagrangian description, unlike theories
with other $T$ values.  $T=1$ string models are the most thoroughly
studied string vacua in the six-dimensional string landscape.
The most widely studied string constructions that give $T=1$ vacua are K3
manifold/orbifold compactifications of the heterotic string 
\cite{GSW6,Strominger,Walton,Sagnotti:1992qw,Erler:1993zy,
smallinstantons,DME,GP,MV,SW6,
Berkooz:1996iz,Honecker}, and
Calabi-Yau threefold compactifications of F-theory 
\cite{MV,AspinwallGross,Sadov,Aldazabal,FMW,Berglund,
Riccioni,Aspinwall}. These were
intensively investigated as they played an essential role in
understanding various string dualities.

The basis most commonly used for $T=1$ theories in the literature is
\be
\label{eq:omega}
\Om=
\begin{pmatrix} 0 & 1 \\ 1 & 0 \end{pmatrix}, \qquad
a=
\begin{pmatrix} -2 \\ -2 \end{pmatrix}, \qquad
b={1 \ov 2 } \lambda_\kappa
\begin{pmatrix} \alpha_\kappa \\ \ta_\kappa \end{pmatrix},\qquad
j={1 \ov \sqrt{2}} \begin{pmatrix} e^{\phi} \\ e^{-\phi} \end{pmatrix}
\ee
for which the factorization condition becomes
\be
\label{genfactorization}
I_8=-{1 \ov 16}(\tr R^2 - \sum_\kappa \alpha_\kappa \tr F_\kappa^2- \sum_{ij} \alpha_{ij} F_i F_j)\wedge
(\tr R^2 - \sum_\kappa \tilde{\alpha}_\kappa \tr F_\kappa^2- \sum_{ij} \tilde{\alpha}_{ij} F_i F_j) \,.
\ee 
For the abelian factors we have
\begin{equation}
b_{ij}={1 \ov 2 }
\begin{pmatrix} \alpha_{ij} \\ \ta_{ij} \end{pmatrix}\,,
\end{equation}
and the gravitational anomaly constraint becomes
\be
H-V = 244\,.
\ee

The kinetic term for the antisymmetric
tensor is given by
\be
\LL = -{1 \ov 2} e^{-2\phi}(dB-\om)\cdot(dB-\om) \,,
\ee
where $\phi$ is the dilaton.
We define the Chern-Simons forms $\om$ and $\tilde{\om}$ as
\begin{align}
d\om &= {1 \ov 16\pi^2} (\tr R^2 - \sum_\kappa \alpha_\kappa \tr F_\kappa^2- \sum_{ij} \alpha_{ij} F_i F_j) \\
d\tilde{\om} &= {1 \ov 16\pi^2} (\tr R^2 - \sum_\kappa \ta_\kappa \tr F_\kappa^2- \sum_{ij} \ta_{ij} F_i F_j)
\end{align}
The variation of the two form under gauge transformations
becomes
\be
\delta B = -{1 \ov 16 \pi^2}
(\sum_\kappa \alpha_\kappa \tr\Lambda_\kappa F_\kappa + \sum_{ij} \alpha_{ij} \Lambda_i F_j)
\ee
and the anomaly can be gotten rid of by adding the term
\be
-B\wedge d\tilde{\om}
\ee
to the Lagrangian. Supersymmetry determines the kinetic term for
the gauge fields to be
\be
-\sum_\kappa (\alpha_\kappa e^\phi +\ta_\kappa e^{-\phi} )\tr F_\kappa\wedge*F_\kappa
-\sum_{ij} (\alpha_{ij} e^\phi +\ta_{ij} e^{-\phi} ) F_i \wedge * F_j \,.
\ee
For a consistent theory without instabilities
there must be a value of the dilaton such that all
the gauge fields have positive kinetic terms. This means that
the matrix
\be
\gamma_{ij} \equiv \alpha_{ij} e^\phi +\ta_{ij} e^{-\phi}
= 2 \sqrt{2} j \cdot b_{ij}
\ee
must be positive definite for some value of $\phi$. Also,
in order for the distinct $U(1)_i$ vector multiplets to be independent
degrees of freedom, $\gamma_{ij}$  must be non-degenerate.

In order to discuss the factorization equations coming from terms
with abelian gauge field factors, in addition to $f_I(x_i) = q_{I,i} x_i$
it is convenient to define the quadratic forms
\be
F(x_i) = \sum_{ij} \alpha_{ij} x_i x_j \qquad \tF(x_i) = \sum_{ij} \ta_{ij} x_i x_j.
\ee
These are the components of $P^\alpha (x)$ defined through \eq{eq:pa}.

The factorization condition can then be summarized by the
polynomial identities
\begin{align}
0 &= \sum_{I} (\MM^\kappa_I E^I_\kappa) f_I(x_i) &\text{for all $\kappa$} \label{pi1}\\
F(x_i) + \tF(x_i) &= {1 \ov 6} \sum_{I} \MM_I f_I(x_i)^2 \label{pi2} \\
\ta_\kappa F(x_i) + \alpha_\kappa \tF(x_i) &= 4 \sum_{I} (\MM^{\kappa}_I A_{\kappa}^I ) f_I(x_i)^2
&\text{for all $\kappa$} \label{pi3}\\
F(x_i) \tF(x_i) &= {2 \ov 3} \sum_I \MM_I f_I(x_i)^4 \label{pi4}
\end{align}

The basis chosen for the $U(1)$ factors is defined up to
$GL(\n,\field{R})$:
\be
\label{glnq}
\begin{pmatrix} \vec{q_1} \\ \vec{q_2} \\ \vdots \\ \vec{q_\n} \end{pmatrix} \rightarrow
M \begin{pmatrix} \vec{q_1} \\ \vec{q_2} \\ \vdots \\ \vec{q_\n} \end{pmatrix},
\quad
(\alpha)_{ij} \rightarrow \left( M^{t}(\alpha) M \right)_{ij},
\quad
(\ta)_{ij} \rightarrow  \left( M^{t}(\ta) M \right)_{ij} \,.
\ee
$(\alpha)$ and $(\ta)$ denote the matrices whose $(i,j)$
element is $\alpha_{ij}$ and $\ta_{ij}$, respectively.

As proven in the last section, since $T=1<9$, the charge
vectors $\{ \vec{q}_i \}$ are linearly independent for solutions
of the factorization equations that give a non-degenerate
kinetic term for some value of the dilaton. Linear independence
of $\vec{q}_i$ imposes positive-definiteness on both
$\alpha_{ij}$ and $\ta_{ij}$. The reason is that the r.h.s.'s of
\eq{pi2} and \eq{pi4} are both positive for any real $x_i$
if $\vec{q}_i$ are linearly independent. This is because
$\vec{f}(x_i)$ cannot be zero for any real $x_i$. Hence
$F(x_i)$ and $\tF(x_i)$ are positive for all real $x_i$.
Therefore, $\alpha_{ij}$ and $\ta_{ij}$ both have to be positive definite.

We now work out two  explicit examples where we can see the equations
at play. The first example is given by orbifold compactifications of
the $E_8 \times E_8$ heterotic string theory \cite{Erler:1993zy}. This theory has
gauge group $E_7 \times E_8 \times U(1)$ with
$10$ $\bold{56}$'s and $66$ singlets with respect to $E_7$.
Nothing is charged under the $E_8$.
This matter structure solves the non-abelian factorization equations.
The non-abelian part of the anomaly polynomial factorizes to
\be
-{1 \ov 16}(\tr R^2 - {1 \ov 6}\tr F_{E_7}^2 - {1 \ov 30}\tr F_{E_8}^2) \wedge
(\tr R^2 - \tr F_{E_7}^2 + {1 \ov 5}\tr F_{E_8}^2) \,.
\ee
We index the hypermultiplet representations $\bold{56}$ by $I=1,\cdots,10$
and the singlets by $I=11,\cdots,76$.
Since there is only one $U(1)$, there is only a single $\alpha = \alpha_{11}$
and a single $\ta=\ta_{11}$. Also, $f_I (x) = q_I x$.

Therefore, the anomaly equations can be obtained by plugging in
\be
F(x)=\alpha x,\quad \tF(x) = \ta x, \quad f_I(x) = q_I x
\ee
to equations \eq{pi1}-\eq{pi4}.
Since $E_7$ and $E_8$ do not have third order invariants, and no
matter is charged under $E_8$, we obtain
\begin{align}
-{1 \ov 5}\alpha+{1 \ov 30} \ta &=0\\
\alpha+ \ta &= {1 \ov 6} (56 \sum_{I=1}^{10} q_I^2 + \sum_{I=11}^{76} q_I^2)\\
\alpha+{1 \ov 6} \ta &= 4 \sum_{I=1}^{10} q_I^2 \\
\alpha\ta &={2 \ov 3} (56 \sum_{I=1}^{10} q_I^4+ \sum_{I=11}^{76} q_I^4)
\end{align}
This can be re-written as
\be
56 \sum_{I=1}^{10} q_I^4+ \sum_{I=11}^{76} q_I^4 = 36 \left( \sum_{I=1}^{10} q_I^2 \right)^2
= {9 \ov 196} \left( \sum_{I=11}^{76} q_I^2 \right)^2 \,.
\ee
Five distinct charge assignments that give solutions
to these equations can be obtained
by different abelian orbifold---by which we mean an orbifold whose
orbifold group is abelian---compactifications.
For example, there is a $Z_8$ orbifold compactification that
assigns the charges
\begin{gather}
q_1=q_2=-3/8, \quad
q_3=q_4=q_5=-1/4, \quad
q_8=q_9=q_{10}=0, \quad
q_6=q_7=-1/8, \nonumber\\ 
q_{11}=\cdots=q_{30}=1/8, \quad
q_{31}=\cdots=q_{34}=-7/8, \quad
q_{35}=q_{44}=1/4, \nonumber \\
q_{45}=\cdots=q_{50}=-3/4, \quad
q_{51}=\cdots=q_{54}=3/8, \quad
q_{55}=\cdots=q_{58}=-5/8,  \nonumber \\
q_{59}=\cdots=q_{77}=1/2, \quad
q_{78}=\cdots=q_{66}=-1/2 \nonumber
\end{gather}
to the hypermultiplets.
The anomaly coefficients for these charge assignments  are
\be
\alpha = 1, \qquad\ta=6.
\ee
All five solutions from abelian orbifolds are
given in table 1 of \cite{Erler:1993zy}.

We present one more example that will prove to be
useful later in this paper.
Consider the gauge group $SU(13)\times U(1)$
with 4 two-index
anti-symmetric, 6 fundamental and 23 singlet representations of $SU(13)$.
These solve the anomaly equations that do not concern the
$U(1)$ field strengths. The non-abelian part factorizes to
\be
-{1 \ov 16} (\tr R^2 - 2\tr F_{SU(13)}^2) \wedge
(\tr R^2 - 2\tr F_{SU(13)}^2 ) \,.
\ee
Denoting the charges of hypermultiplets in the
antisymmetric/fundamental/singlet representations as
$a_x$($x=1,\cdots,4$)/$f_y$($y=1,\cdots,6$)/$s_z$($z=1,\cdots,23$)
the anomaly equations become
\begin{align}
0&= \sum_x 9 a_x + \sum_y f_y\\
\alpha + \ta &= {1 \ov 6} ( \sum_x 78 a_x^2 + \sum_y 13 f_y^2 + \sum_z s_z^2 )\\
2\alpha + 2\ta &=4( \sum_x 11 a_x^2 + \sum_y f_y^2 ) \\
\alpha \ta &= {2 \ov 3} (\sum_x 78 a_x^4 + \sum_y 13 f_y^4 + \sum_z s_z^4)
\end{align}
If there exist for given $a_x, f_y, s_z$ a solution $\alpha, \ta$
to these equations, the anomaly polynomial factorizes into
\be
-{1 \ov 16}(\tr R^2 - 2\tr F_{SU(13)}^2 -\alpha F_{U(1)}^2) \wedge
(\tr R^2 - 2\tr F_{SU(13)}^2 -\ta F_{U(1)}^2)
\ee
We identify infinite classes of charge assignments
and $\alpha,\ta$ values that solve these equations in section \ref{s:inf}.

\subsection{$T=0$} \label{ss:teq0}

We now discuss anomaly cancellation
and the factorization equations in the case of $T=0$.
Theories without tensor multiplets are in some ways the simplest  type of 6D
supergravity theory.  Although such vacua do not arise directly from
geometric heterotic or type II compactifications, $T = 0$ vacua are easily
constructed in F-theory by compactifications on the base $\field{P}^2$, and
can be reached from $T = 1$ vacua by tensionless string transitions
\cite{MV, SW6}.
F-theory models for $T = 0$ vacua based on toric Calabi-Yau threefolds
that are elliptic fibrations over $\field{P}^2$ have recently been
systematically studied in \cite{Braun}.
Analysis of 6D supergravity theories with $T = 0$ 
are given in \cite{Kumar:2010am}, and F-theory vacua giving $T = 0$
vacua are systematically studied in
\cite{Morrison-Taylor}.
Further references for string constructions of
$T=0$ vacua can be found in these papers.

In the case $T = 0$ all the $SO(1,T)$ vectors  $a, b, j$ reduce to
numbers. $j^2=1$ leaves us with a sign ambiguity.
Without loss of generality we can set $j=1$.
Positivity of the kinetic term imposes that the $b_\kappa$'s be
positive and that $b_{ij}$ be a positive definite matrix.
This and equation (\ref{1st}) sets $a<0$.
The relation $a^2 =9-T$ fixes $a=-3$.
To summarize,

\be
\Om=1, \qquad
a=-3, \qquad
j=1 \,.
\ee

In this case the factorization condition becomes
\be
\label{genfactorizationt0}
I_8=-{1 \ov 32} (-{3 \ov 2}\tr R^2
+ \sum_\kappa {2 b_\kappa \ov \lambda_\kappa} \tr F_\kappa^2+ \sum_{ij} 2b_{ij} F_i F_j)^2 \,.
\ee
The gravitational anomaly constraint becomes
\be
H-V = 273 \,.
\ee

The factorization equations coming from $U(1)$'s
can be written out by using the quadratic form
\be
P(x_i) = \sum_{ij} b_{ij} x_i x_j \,,
\ee
as the polynomial identities
\begin{align}
0 &= \sum_{I} (\MM^\kappa_I E^I_\kappa) f_I(x_i) &\text{for all $\kappa$} \label{pi10}\\
P(x_i) &= {1 \ov 18} \sum_{I} \MM_I f_I(x_i)^2 \label{pi20} \\
P(x_i) &= { \lambda_\kappa \ov b_\kappa} \sum_{I} (\MM^{\kappa}_I A_{\kappa}^I ) f_I(x_i)^2
&\text{for all $\kappa$} \label{pi30}\\
P(x_i)^2 &= {1 \ov 3} \sum_I \MM_I f_I(x_i)^4 \label{pi40}
\end{align}

The basis chosen for the $U(1)$ factors is as usual defined up to
$GL(\n,\field{R})$ through \eq{eq:invariance}.
Since $T=0<9$, the charge vectors $\{ \vec{q}_i \}$
are linearly independent for solutions
of the factorization equations that give a non-degenerate
kinetic term for some value of the dilaton.

As in the $T=1$ case it is useful to look at a few examples in which
the anomaly equations come into play.  We first consider a theory with
gauge group $SU(6)\times U(1)$ with 1 adjoint, 9 two-index
anti-symmetric, 18 fundamental and 31 singlet representations of
$SU(6)$. These solve the non-abelian anomaly factorization
equations. The factorized non-abelian anomaly polynomial is
\be
-{1 \ov 32}(-{3 \ov 2}\tr R^2 + 6 \tr F_{SU(6)}^2)^2 \,.
\ee
Denoting the charge of hypermultiplets in the
adjoint/antisymmetric/fundamental/singlet representation as
$d$/$a_x(x=1,\cdots,9)$/$f_y (y=1,\cdots,9)$/$s_z(z=1,\cdots,31)$ the
anomaly equations become
\begin{align}
0&= \sum_x 5 a_x + \sum_y f_y \label{eq:0-1}\\
b &= {1 \ov 18} (35 d^2+ \sum_x 15 a_x^2 + \sum_y 6 f_y^2 + \sum_z s_z^2 )\\
3b &= ( 12 d^2 + \sum_x 4 a_x^2 + \sum_y f_y^2 ) \\
b^2 &= {1 \ov 3} (35d^4 +\sum_x 15 a_x^4 + \sum_y 6 f_y^4 + \sum_z s_z^4)
\label{eq:0-4}
\end{align}
For charges and $b$ satisfying these equations, the anomaly polynomial
of the theory factorizes into
\be
-{1 \ov 32}(-{3 \ov 2}\tr R^2 + 6 \tr F_{SU(6)}^2 + 2b F_{U(1)}^2)^2 \,.
\ee
Finding an apparently consistent supergravity theory with this gauge
group amounts to identifying values for $b$ and the charges $d, a_x,
f_y, s_z$ so that \eq{eq:0-1} through \eq{eq:0-4} are satisfied.  If
we assume that the $U(1)$ is compact and the charges are integers then
this is a system of Diophantine equations over the integers.  In
general, classifying solutions to such a system of equations can be a
highly nontrivial problem in number theory.

A particularly interesting class of examples are pure abelian theories.
In this case the only non-trivial abelian anomaly equations are
equations \eq{one} and \eq{four} (\eq{eq:0-1} and \eq{eq:0-4} for $T =
0$).
For a theory with a given number of abelian vector multiplets,
there is a lower bound on the number of charged multiplets it must have.
When the number of charged hypermultiplets saturate this bound,
the charges that the hypermultiplets carry is severely restricted.
Such theories have a particularly simple structure and are interesting
to study further.

As an example,  consider the case of a purely abelian theory when $T=0$ and $V_A=1$.
We denote the charges of the $X$ charged hypermultiplets in the
theory by $q_1,\cdots,q_X \neq 0$.
Then we must solve
\begin{align}
\begin{split}
18b &= \sum_I q_I^2, \\
3b^2 &= \sum_I q_I^4.
\end{split}
\label{hchargeex}
\end{align}
Using the inequality
\be
(\sum_I q_I^2)^2 \leq X (\sum_I q_I^4),
\ee
we see that
\be
X \geq 108.
\ee
When $X$ is equal to $108$,  {\it i.e.,} when the number of charged
hypermultiplets saturates the lower bound, the only solutions
to the equations \eq{hchargeex} are
\be
q_I = \pm Q\quad \text{for all $I$}, \quad b = 6 Q^2.
\ee

Similarly for any pure abelian $T=0$ theory,
using \eq{pi10} and \eq{pi40} we can show
that the following relation between $V_A$
and the number of charged
hypermultiplets $X$ holds
\be
{324 \, V_A \ov V_A +2} \leq X \leq V_A + 273.
\label{purebound}
\ee
The proof is given in appendix \ref{ap:purebound}.
Hence, as above, when $V_A=1$ there must be at least 108 charged
hypermultiplets; likewise, when $V_A=2$ there must be
at least 162 charged hypermultiplets.
As seen in the $V_A=1$ case,
in the marginal cases when $X$ exactly saturates this bound,
the solutions to the charge equations are particularly simple.
From (\ref{purebound}) it follows that the maximum possible number of
$U(1)$ factors that can be included in any $T = 0$ theory with
no nonabelian gauge group is $V_A \leq 17$.\footnote{(\ref{purebound})
alone implies that $V_A \leq 17$ or $V_A \geq 32$.
An additional constraint
following from equations \eq{pi10} and \eq{pi40} is needed to
obtain the desired bound. We derive this constraint and show
that indeed $V_A \leq 17$ in appendix \ref{ap:purebound}.}

A family of marginal/nearly marginal
$T=0$  theories with gauge group $U(1)^k, k \leq 7$ can be obtained by Higgsing an $SU(8)$ theory with
one adjoint hypermultiplet and nine antisymmetric hypermultiplets.
The number of charged hypermultiplets $X$ for the
various pure abelian theories one obtains by Higgsing the
adjoint of this theory in different ways is summarized in table \ref{t:pureab}.
An F-theory construction of this $SU(8)$ model, 
which has $b = 3$,
through an explicit
Weierstrass model is described in \cite{Morrison-Taylor}.  In
principle the adjoint in this construction can be Higgsed to give an
F-theory description of the full family of marginal/near marginal
$U(1)^k$ models, though we have not worked out the details of this Higgsing.

\begin{table}[t!]
\center
 \begin{tabular}{ | c| c | c | c | c | c | c | c | c | }
 \hline
 Gauge Group & $\cdot$ & $U(1)$ & $U(1)^2$ & $U(1)^3$ & $U(1)^4$ & $U(1)^5$ & $U(1)^6$ & $U(1)^7$  \\ \hline
 $X$ & 0 & 108 & 162& 198& 225 & 243 & 252 & 252 \\ \hline
 $324\,V_A/(V_A+2)$ & 0 & 108 & 162 & 194.4... & 216 & 231.4... & 243 & 252 \\ \hline
 $X'$ & 273 & 166 & 113 & 78 & 52 & 35 & 27 & 28 \\ \hline
  \end{tabular}
 \caption{The number of charged hypermultiplets $X$ for
pure abelian theories obtained by Higgsing the
adjoint of the $SU(8)$ theory with one adjoint and nine antisymmetrics.
We have also tabulated the number of uncharged hypermultiplets
in the theory, $X'=(273+V_A-X)$.}
\label{t:pureab}
\end{table}

Explicit F-theory compactifications are known for the first four
theories on this table. Six-dimensional $\NN=(1,0)$ theories with $T=0$
can be obtained by F-theory compactifications on Calabi-Yau threefolds
that are elliptic fibrations of $\mathbb{P}^2$ \cite{MV}.
Such a Calabi-Yau threefold that is non-singular can be expressed as
a degree 18 hypersurface in the projective space
$\mathbb{P}[1,1,1,6,9]$ \cite{KlemmLianRoanYau}, which is denoted by
\be
X_{18}[1,1,1,6,9]^{2,272}.
\ee
The subscript denotes the degree of the hypersurface, the
number in the brackets parametrize the projective space,
and the two superscripts denote the $h_{1,1}$ and $h_{2,1}$
values of the manifold.
For $T=0$ vacua, the total rank of the gauge group is
given by $(h_{1,1}-2)$ and the number of uncharged hypermultiplet
is given by $(h_{2,1}+1)$ \cite{MV}. It is easy to check that the
data of this manifold reproduces the first theory in table \ref{t:pureab}.

There is a general process by which one can replace the fiber-type of
an elliptically fibered manifold to generate a different manifold
\cite{KlemmLianRoanYau}.
From the point of view of stringy geometry, one can understand this
as a conifold transition between topologically distinct manifolds \cite{StromingerConifold}.
Three manifolds can be generated from $X_{18}[1,1,1,6,9]^{2,272}$
by successive conifold transitions.
They are given by
\be
X_{12}[1,1,1,3,6]^{3,165}, \qquad
X_9[1,1,1,3,3]^{4,112}, \qquad
X_{6,6}[1,1,1,3,3,3]^{5,77}. \label{man}
\ee
At a generic point in the complex moduli space, theories obtained
by compactifying on these manifolds do not have nonabelian gauge symmetry.
Comparing the numbers with table \ref{t:pureab},
we find that the massless spectrum of the six-dimensional theories obtained
by F-theory compactifications on the three manifolds of \eq{man} coincides with
the massless spectrum of the second, third and fourth theories of
table \ref{t:pureab} with gauge groups $U(1)^k, k = 1, 2, 3$. 
We do not know how to continue this process to construct an explicit
geometry realizing a theory with gauge group $U(1)^4$.

\subsection{Linear Multiplets and Generalized Green-Schwarz Anomaly Cancellation}
\label{ss:Stuckelberg}

In this section we discuss linear multiplets and their role in the
generalized Green-Schwarz anomaly cancellation mechanism.  We first
discuss how two different types of hypermultiplets can be
distinguished when we consider their representation under $SU(2)_R$.
Then we show how each multiplet couples to
vector multiplets.  In particular, we show how a linear multiplet can
couple to an abelian vector multiplet and form a long multiplet.  Next
we depict the role that linear multiplets play in the generalized
Green-Schwarz anomaly cancellation mechanism.  Lastly we show that we
may ignore long multiplets formed in this way and the generalized
Green-Schwarz mechanism when we are discussing the massless spectrum
of the theory.  Most of the information on linear multiplets given in
this section can be found in \cite{Douglas:1996sw}.

There are two different kinds of hypermultiplets in supersymmetric 6D
theories with 8 supercharges.
The scalar components of the hypermultiplet
can transform either as a complex $\bold{2}$ or a real
$\bold{3} + \bold{1}$ under the $SU(2)_R$ symmetry of the theory.
We refer to the first type of hypermultiplet simply as hypermultiplets,
and the second kind of hypermultiplet as linear multiplets. As far
as their contribution to the gravitational anomaly are concerned,
the two kinds of hypermultiplets behave identically.
The fermions of the linear multiplet are not charged under
any gauge group, so the contribution of the linear multiplet to the anomaly is
equivalent to that of a neutral hypermultiplet, as shown shortly.


As stated above, under the $SU(2)_R$ symmetry,
the scalar components of the hypermultiplet transform as a complex
$\bold{2}$. The spinors, on the other hand are neutral, {\it i.e.,} 
singlets ($\bold{1}$). 
Meanwhile, the scalar components of the linear multiplet transform as
a real $\bold{3}+\bold{1}$. The spinors transform as $\bold{2}$'s. 

To see how these multiplets couple to other fields, it is useful to
reduce to four dimensions on a two-torus and write out the Lagrangian
in terms of $\NN=1$ superfields. Both multiplets, when dimensionally
reduced, are $\NN=2$ fields that consist of two chiral superfields.
The hypermultiplets can transform in a non-trivial representation of the
gauge group and consist of two chiral superfields $Q$ and $\tilde{Q}$.
In this case, the representation of $Q$ must be the conjugate of
that of $\tilde{Q}$. It is well known that this multiplet couples to the $\NN=2$
vector multiplet that consists of a vector multiplet $V$ and a chiral
multiplet $\Phi$ in the adjoint representation as
\be
\int d^4 x d^4 \th \; ( Q^\dagger e^V Q + \tilde{Q}^\dagger e^{-V}\tilde{Q} )
+ \int d^4 x d^2 \th \; \tilde{Q}^T \Phi Q + (h.c.) \,.
\ee

Meanwhile, the linear multiplets couple to other fields in quite a
different manner \cite{Douglas:1996sw}. They cannot couple to gauge
fields in the standard way, as the Lagrangian would not be $SU(2)_R$
invariant in this case. They can couple to $U(1)$ gauge fields, however.
The linear multiplet consists of two chiral fields $C$ and $B$ and
couples to $U(1)$ gauge fields as
\be
\int d^4x d^4 \th \; ({1 \ov 2}(iC-iC^\dagger-V)^2 + B^\dagger B)
- {1 \ov \sqrt{2}}\int d^4x d^2 \th \; B \Phi + (h.c.) \,.
\ee
Writing the scalar of $C$ as $(\pi_3+i\phi)$ and the scalar of $B$ as
$(\pi_1+i\pi_2)$, the kinetic terms for the scalars become
\be
-\int d^4 x ( (\p_\mu \phi - {1 \ov 2}A_\mu)^2 + (\p_\mu \pi_i )^2)  \,.
\ee
The $\phi$ can be gauged away
using the gauge transformation
\be
A_\mu \rightarrow A_\mu + \p_\mu \Lambda,
\qquad \phi\rightarrow\phi+{1 \ov 2} \Lambda \,,
\ee
and the $U(1)$ gauge field obtains mass $1/2$. 
The $U(1)$ gauge field has recived a mass
by the St\"uckelberg mechanism.

By integrating out the F-terms of the linear multiplet,
we see that the scalar in $\Phi$ recieves the same mass($1/2$).
Meanwhile, the fermions do not couple to the gauge field,
and hence only contribute to gravitational anomalies.
They only couple to the fermions in $V$ and $\Phi$
through Dirac mass terms, {\it i.e.,} fermions of $C$ and $B$
pair up with fermions of $V$ and $\Phi$ into two Dirac fermions
of mass $1/2$.

Combining the auxiliary fields of $V$ and $\Phi$, we get
three real auxiliary fields (the `D fields' for the $\NN=2$ vector multiplet)
that are in the $\bold{3}$ of $SU(2)_R$. These couple to the scalars
transforming as the $\bold{3}$:
\be
-\int d^4 x (\pi_i D^i )
\ee
Expanding around a vacuum with $\pi_i=0$,
the $U(1)$ vector multiplet and linear multiplet together form
a long $\NN=2$ multiplet with 5 scalars, 2 Dirac fermions,
and a vector field, all of mass $1/2$ in units of the mass parameter.
Note that this long massive spin-1 multiplet is not chiral,
as the fermions are Dirac.

When we have linear multiplets, they may be used to cancel
anomalies. As discussed in section \ref{s:comments},
it is possible to cancel anomalies of the form
\be
F_i \wedge X_6 \,,
\ee
where $X_6$ is a six form, by adding the term
\be
-\phi \wedge X_6 \,.
\ee
$\phi$ is a St\"uckelberg 0-form inside a linear multiplet.

In order for the generalized anomaly
cancellation to work, we must have a linear multiplet at our disposal.
If we do not have such a linear multiplet, we cannot get rid of the
term and hence the theory would be anomalous. In case we
have such a multiplet, through the St\"uckelberg mechanism, we expect
the linear multiplet to be eaten to form a long massive
spin-1 multiplet. Schematically, we may write
\be
L_i = V_i + H_i \,,
\ee
where $L_i$ denotes the long multiplet, $V_i$ denotes the $U(1)$
vector multiplet, and $H_i$ the linear multiplet.

So we see that all the vector bosons of $U(1)$ gauge symmetries
whose anomalies are cancelled in this fashion must be
massive and must form a long multiplet.
These long multiplets are non-chiral and hence do not contribute
to gravitational anomalies. Furthermore none of the fields inside
this multiplet are charged under other gauge groups.
Therefore, we see that these multiplets contribute neither to
gravitational anomalies nor to unbroken gauge/mixed anomalies.

By this logic we can further state that all long multiplets
obtained by $U(1)$ gauge bosons coupling to linear multiplets
do not contribute to the anomaly polynomial. Therefore, we may ignore all
the long multiplets---or vector/linear multiplet pairs that couple---when we
are discussing gravitational anomalies and gauge/mixed anomalies
concerning unbroken gauge symmetry,
{\it i.e.,} gauge symmetry of the massless spectrum.

Long multiplets and hence the generalized Green-Schwarz
mechanism may thus be ignored when we are discussing the massless
spectrum of the theory. In other words, when we are constructing
low-energy effective theories, writing down anomalous $U(1)$'s and
then lifting them is a redundant procedure.
We may safely restrict our attention to the massless spectrum
whose anomalies are all cancelled by two-forms;
the factorization condition (\ref{genfactorization0}) should hold
for these theories.

\section{Bounds on $T < 9$ Theories With $U(1)$'s} \label{s:fin}

We now address the first (I) of the four questions raised in the introduction.
That is, we prove 
that the number of different gauge/matter structures--- specified
by the gauge group and the non-abelian representation of the matter---is
finite for theories with $T < 9$, when we ignore the charge of the
matter under the $U(1)$'s. 

The strategy we pursue is the following.
First in section \ref{ss:boundU1}, we prove that in a non-anomalous
theory, the number of $U(1)$'s is bounded by a number
determined by the non-abelian gauge/matter content.
We prove that the relations
\begin{align}
\n &\leq (T+2)\sqrt{2N}+2(T+2) \label{centone} \\
\n &\leq (T+2)(T + {7 \ov 2}) + (T+2) \sqrt{2 V_{NA} + (T^2-51T+{2225 \ov 4})} \label{centtwo} 
\end{align}
hold for non-anomalous theories with $T<9$, where $\n$ is the rank of the
abelian gauge group, $V_{NA}$ is the number of nonabelian vector multiplets,
and $N$ is the number of hypermultiplet representations.
These bounds imply that the number of $U(1)$'s one could
add to a non-abelian theory is finite.
We note that these bounds are in no sense optimal; they could be
improved by a more careful analysis.
These inequalities, however, will be sufficient for the purpose
of proving that there is a finite bound on theories with $T<9$.

In section \ref{ss:curability} we define the concept of `curable theories'
as non-abelian theories with $H-V >273-29T$ that can be made
non-anomalous by adding $U(1)$ vector fields and without changing
the non-abelian gauge/matter structure.
Curable theories are defined so that all non-anomalous
theories with abelian gauge symmetry can be obtained  by adding
$U(1)$'s either to non-anomalous theories, or to curable theories.
We then show that the number of  curable theories is finite for $T <
9$, which combined with our other results  
implies that the number of gauge/matter structures possible for
non-anomalous theories with $T < 9$ is finite.

In section \ref{ss:Tgeq9} we construct an infinite class of non-anomalous
theories with an unbounded number of $U(1)$'s and $T \geq 9$.

\subsection{Bound on Number of $U(1)$ Factors} \label{ss:boundU1}

In this section we prove equations \eq{centone} and \eq{centtwo}
for non-anomalous theories
with $T<9$. Given a gauge group
\be
\GG = \prod_{\kappa=1}^\nu \GG_\kappa \times \prod_{i=1}^\n U(1)_i \,,
\ee
we show that the bound  on $\n$ can be given as a function of
the number of nonabelian vector multiplets 
\be
V_{NA}=\sum_\kappa d_{\text{Adj}_\kappa}
\ee
and $N$, the number of nonabelian matter representations.

This can be done by making use of equation (\ref{4th}),
which is  equivalent to \eq{four} :
\begin{align}
\begin{split}
P(x_i) \cdot P(x_i) &={1 \ov 3} \sum_I \MM_I f_I(x_i)^4 \\
\Leftrightarrow
b_{ij} \cdot b_{kl} + b_{ik} \cdot b_{jl} +b_{il} \cdot b_{jk}
&= \sum_I \MM_I q_{I,i}q_{I,j}q_{I,k}q_{I,l}
\label{4ag}
\end{split}
\end{align}

We should be looking for integral solutions of this equation for
$b_{ij}, q_{I, i}$,
but for now we simply determine the conditions for the
equations to have real solutions. These conditions impose a bound
on $\n$, which also is a bound for integral solutions.
These equations have a $GL(n,\RR)$ invariance summarized by
\eq{glnq} where the matrices $M$ now can be taken to be real.

We first state the following useful

\medskip

\noindent\textbf{Fact :} For $(T+1)$ symmetric $n \times n$ matrices
$S_1, \cdots, S_{T+1}$, there exists a matrix $M \in GL(n,\field{R})$ such
that for $\tau \equiv \lceil n/{(T+2)} \rceil$ the matrices $S'_\alpha = M^t S_\alpha M$
satisfy
\be
(S'_\alpha)_{kl} = 0 \qquad \text{for distinct }k, l \leq \tau
\ee
for all $\alpha=1,\cdots,(T+1)$.

\medskip

\noindent\textbf{Proof :} First pick an arbitrary $n$-dimensional
vector $e_1$. Then generate the set of $(T+2)$ vectors
\be
V_1 = \{ e_1, S_1 e_1, \cdots, S_{(T+1)} e_1 \} \,.
\ee
When $1 < n/(T+2)$ there always exists a non-zero
vector that is orthogonal to these $(T+2)$ vectors. Pick one and call it $e_2$.
Then generate the set of $(T+2)$ vectors
\be
V_2 = \{ e_2, S_1 e_2, \cdots, S_{(T+1)} e_2 \} \,.
\ee
When $2 < n/(T+2)$ there always exists a non-zero
vector that is orthogonal to the set $V_1 \cup V_2$ of vectors.
Pick one and call it $e_3$. By iterating this process we can obtain $\tau$
non-zero mutually orthogonal vectors,
\be
e_1,\cdots, e_\tau
\ee
such that
\be
e_i^t S_\alpha e_j =0 \quad \text{for } i \neq j
\ee
for all $\alpha$. We can then choose vectors $e_{\tau+1}, \cdots, e_{n}$
that together with $e_1,\cdots e_\tau$ form a basis of $\field{R}^n$.
Define
\be
M=(e_1 \cdots e_n) \,,
\ee
where $e_i$ are column vectors.
It is clear that $\det M \neq 0$ and that for $S'_\alpha = M^t S_\alpha M$
\be
(S'_\alpha)_{kl} = 0 \qquad \text{for distinct }k, l \leq \tau
\ee
for all $\alpha$. $\Box$

\medskip

Due to this fact, there exists a matrix $M \in GL(\n,\field{R})$
such that for all $\alpha$
\be
M_{ik} M_{jl} b^\alpha_{kl}=0 \quad \text{for distinct }i, j\leq \tau
\ee
for any solution of \eq{4ag}. We have defined
\be
\tau = \left\lceil {\n \ov T+2} \right\rceil \,.
\ee
This means that the existence of a solution of \eq{4ag}
implies the existence of a solution of the same equations with
\be
\vec{b}_{kl}=0 \quad \text{for distinct }k,l\leq \tau \,.
\ee
Therefore, we may from now on assume that this condition
is true.

For ordered pairs $(i,j)$ with $i<j \leq \tau$, we define the vectors
\be
\vec{Q}_{ij} \equiv (\sqrt{\MM_1}q_{1,i}q_{1,j} ,
\sqrt{\MM_2}q_{2,i}q_{2,j},\cdots,\sqrt{\MM_N}q_{N,i}q_{N,j}) \,.
\ee
Then we have
\be
\vec{Q}_{ij} \cdot \vec{Q}_{kl}= \sum_I \MM_I q_{I,i}q_{I,j}q_{I,k}q_{I,l}
=b_{ij} \cdot b_{kl} + b_{ik} \cdot b_{jl} +b_{il} \cdot b_{jk}=0
\ee
for ordered pairs $(i,j) \neq (k,l)$.
Also from equation (\ref{four}), we have
\be
\vec{Q}_{ij} \cdot \vec{Q}_{ij} =\sum_I \MM_I q_{I,i}^2q_{I,j}^2
=b_{ii} \cdot b_{jj} + 2b_{ij} \cdot b_{ij} =b_{ii} \cdot b_{jj} >0. \label{Qnon0}
\ee
The last inequality holds due to the fact that $\vec{b}_{ii}$ are
timelike vectors since
\be
|\vec{b}_{ii}|^2 = {1 \ov 3} \sum_I \MM_I q_{I,i}^4 >0 \,,
\ee
as $\vec{q}_i$ cannot be a zero vector.
The inner-product of two timelike $SO(1,T)$ vectors cannot be zero,
and $\sum_I \MM_I q_{I,i}^2q_{I,j}^2$ cannot be negative and
hence the inequality in \eq{Qnon0}.

Therefore, $\vec{Q}_{ij}$ are non-zero mutually orthogonal vectors for $i < j \leq \tau$.
Thus, we have $\tau(\tau-1)/2$ non-zero orthogonal vectors
in an $N$-dimensional space. Hence
\be
{1 \ov 2}({\n \ov T+2}-1)({\n \ov T+2}-2) \leq {\tau(\tau-1) \ov 2} \leq N \leq H \leq V_{NA}+\n+273-29T.
\ee
Using the two inequalities
\begin{align}
{1 \ov 2}({\n \ov T+2}-1)({\n \ov T+2}-2) &\leq N \\
{1 \ov 2}({\n \ov T+2}-1)({\n \ov T+2}-2) &\leq V_{NA}+\n+273-29T,
\label{eq:second}
\end{align}
we obtain the bounds
\begin{align}
\n &\leq (T+2)\sqrt{2N}+2(T+2)\label{centralprime0} \\
\n &\leq (T+2)(T + {7 \ov 2}) + (T+2) \sqrt{2 V_{NA} + (T^2-51T+{2225 \ov 4})} \,,\label{centralprime}
\end{align}
as promised.
We have used the fact that $\sqrt{a+b} \leq \sqrt{a} +\sqrt{b}$
for non-negative $a,b$ to simplify the first inequality.
The second inequality simply follows from solving 
(\ref{eq:second}) for $V_A$ when the inequality is saturated.
This result implies that given a non-anomalous non-abelian theory,
the number of $U(1)$'s one could add to the theory keeping it
non-anomalous is bounded.


The equations we have used also apply to pure abelian
theories. This is because we have not used any constraint coming
from the non-abelian structure of the theory; we have only used
the equation (\ref{4ag}).  Hence we can obtain a
bound on the number of $U(1)$'s when the theory is purely abelian:
\be
\label{roughbound}
\n \leq (T+2)(T + {7 \ov 2}) + (T+2) \sqrt{T^2-51T+{2225 \ov 4}}
\ee
Note that this bound is substantially weaker than the tighter bound $\n \leq 17$
for $T = 0$. (For $T = 0$
this bound states that $\n \leq  54$, while we show that $\n \leq 17$
in appendix \ref{ap:purebound}.)

\subsection{Curability and Finiteness of Curable Theories} \label{ss:curability}

We define `curable' theories  to be
non-abelian theories that violate the gravitational anomaly bound $H-V >273-29T$, but
whose anomaly polynomial can nonetheless be
made factorizable by adding $U(1)$ vector multiplets
and some singlet hypermultiplets in such a way that the gravitational bound is satisfied.
There should also exist 
values for the scalars in the tensor multiplets
that make the kinetic terms of all gauge fields positive
in the resulting non-anomalous theory.
We also assume that these theories do not have any hypermultiplets
that are singlets under the non-abelian gauge group.

From this definition it is clear that all non-anomalous theories
with abelian gauge symmetry can be obtained by the following steps.
\begin{enumerate}
\item Begin with a theory without abelian gauge group factors that is either non-anomalous or curable.
\item Add abelian vector multiplets and (possibly) hypermultiplets in the
trivial representation of the non-abelian gauge group.
\item Assign $U(1)$ charges to the matter.
\end{enumerate}
We note that it is clear that the number of $U(1)$'s one could add
to a given curable theory is finite, since it is bounded by \eq{centralprime0}
and \eq{centralprime}.
From this it is evident that the crucial remaining step in obtaining bounds
on theories with abelian gauge symmetry
is showing that the number of curable theories is
bounded.

As an example of a curable theory,  consider
the $T=0$ theory with gauge group and matter content
\begin{equation}
SU(9): \;\;\;\;\; \;\;\;\;\;
26 \times {\tiny\yng(1)}+
1 \times {\tiny\yng(1,1)}+
1 \times {\tiny\yng(1,1,1)},
\;\;\;\;\; (H-V = 274) \,.
\label{eq:curable-example}
\end{equation}
Although this theory violates the gravitational anomaly bound,
it satisfies the other gauge/mixed anomaly equations with $b=2$.
(Note that, as explained in section \ref{ss:teq0}, when $T=0$
the anomaly coefficients $a, b$ are numbers.)
The theory (\ref{eq:curable-example}) can be cured by adding a single $U(1)$
vector multiplet and assigning 
charges to the matter in the following way
\begin{align}
\begin{split}
SU(9) \times U(1): \;\;\;\;\;
&6 \times ({\tiny\yng(1)}\,,+1)+
6 \times ({\tiny\yng(1)}\,,-1)+
14 \times ({\tiny\yng(1)}\,,\cdot)+ \\
&1 \times ({\tiny\yng(1,1)}\,,\cdot)+
1 \times ({\tiny\yng(1,1,1)}\,,\cdot),
\qquad (H-V = 273) \,.
\end{split}
\end{align}
The anomaly polynomial of the final theory factorizes to
\be
I_8 = -{1 \ov 32} (-{3 \ov 2} \tr R^2 + 2 \tr F^2_{SU(9)} + 6 F_{U(1)}^2)^2 \,.
\ee
A systematic classification of $T = 0$ supergravity models with
$SU(N)$ gauge groups and without
anomalies or other inconsistencies was given in
 \cite{Kumar:2010am}, and F-theory constructions of such models were
 analyzed in\cite{Morrison-Taylor}.
The
methods and results of those papers, in which
abelian gauge group factors were not treated, can be expanded to
include curable models such as this  $SU(9)$  theory.  In particular,
this presents a particularly simple example of a model with a 3-index
antisymmetric representation of $SU(9)$
for which an F-theory realization might be constructed.

Since abelian vector multiplets and singlet hypermultiplets
do not appear in a curable theory and do not contribute to the nonabelian gauge/mixed anomalies,
it is clear that the anomaly polynomial of a curable theory
takes the form
\begin{align}
\begin{split}
I_8 =
&-{(H-V-273+29T) \ov 5760} \left( \tr R^4 + {5 \ov 4} (\tr R^2 )^2 \right) \\
&-{1 \ov 32} \Om_{\alpha\beta}
\left( {1 \ov 2} a^\alpha \tr R^2 +\sum_\kappa ({2b_\kappa^\alpha \ov \lambda_\kappa}) \tr F_\kappa^2 \right)
\left( {1 \ov 2} a^\beta \tr R^2 +\sum_\kappa ({2b_\kappa^\beta \ov
  \lambda_\kappa}) \tr F_\kappa^2 \right) \,, \label{curablepoly}
\end{split}
\end{align}
where $H-V$ is larger than $273-29T$.
One might think that any  theory of this type is naively curable,
since we could apparently add an arbitrary number
of $U(1)$ vector multiplets under which no matter field is charged, so
that $H-V'=273-29T$.
The kinetic term for these vector fields, however, would
be degenerate---in fact zero---if we do so.
In fact, in many cases the bounds
on the number of $U(1)$ factors that can be
added to a theory make it impossible
to cure nonabelian theories with anomalies
of the form \eq{curablepoly}.


In \cite{Kumar:2009ae, Kumar:2010ru} it was proven that the number of
distinct nonabelian gauge groups and matter representations possible
for theories with $T < 9$ and no $U(1)$ factors is finite.  The bound
$H_{\rm charged}  -V\leq 273-29T$ from the gravitational anomaly
condition played a key role in this proof, limiting the number of
charged hypermultiplets that could appear in a theory with any given
nonabelian gauge group.  To prove that the number of curable theories is also
finite for $T < 9$ we need an analogous constraint on the
number of hypermultiplets for theories with $U(1)$ factors.
We now find such a bound, using the bounds
 \eq{centralprime0}
and \eq{centralprime} on the number of $U(1)$ factors that can be
added to a curable theory.

Suppose a theory is curable by adding $\n$ $U(1)$ vector multiplets
and $H'$ hypermultiplets. Then using \eq{centralprime0} we obtain
\begin{align}
\begin{split}
273-29T &\geq (H-V)_\text{cured theory}
=H-V+H'-\n \\
&\geq H-V+H'-\sqrt{2}(T+2)\sqrt{H'+N}-2(T+2) \\
&= (H-V)+(\sqrt{H'+N}-{(T+2)\ov \sqrt{2}})^2-N-({1\ov 2} T^2 +4T+6) \,,
\end{split}
\end{align}
where $H$, $V$ and $N$ denote the numbers of hypermultiplets,
vector multiplets and hypermultiplet representations in the initial
non-abelian theory.
Since $H'\geq 0$, when $N \leq (T+2)^2/2$ we have
\be
 (\sqrt{H'+N}-{(T+2)\ov \sqrt{2}})^2-N \geq -{(T+2)^2 \ov 2} \,,
\ee
while when $N \geq (T+2)^2/2$ we have
\be
 (\sqrt{H'+N}-{(T+2)\ov \sqrt{2}})^2-N \geq
(\sqrt{N}-{(T+2)\ov \sqrt{2}})^2-N=-\sqrt{2}(T+2)\sqrt{N}+{(T+2)^2 \ov 2} \,.
\ee

Thus, any curable theory
satisfies one of the following two constraints:
\begin{align}
H-{V} &\leq 273-29T + (T^2+6T +8) \label{modcon1} \\
H-{V}-\sqrt{2}(T+2) \sqrt{N} &\leq 273-29T + (2T+4) \label{modcon2}
\end{align}
Curable theories therefore must satisfy the non-abelian factorization equations
\eq{nab}--\eq{nae} and one of these modified gravitational anomaly constraints.

This result suggests that 
the proof in \cite{Kumar:2009ae, Kumar:2010ru}
can be modified to show that the number of curable theories are in fact finite.
There it was shown that the $H$ of theories that obey
the non-abelian factorization equations---and
can have a positive kinetic term---grew faster than the $V$
of the theory when $V$ became large.
This in turn implied that $V$ must be bounded for theories that satisfy the
the non-abelian factorization equations and respect the $H-V$ bound.
We have shown that curable theories must obey the same
non-abelian factorization equations with the $H-V$ constraint modified.
Fortunately, this constant is only modified by a term subleading in $N<H$.
This suggests
that the boundedness of
curable theories can be shown along the same lines as the proof of
boundedness of non-abelian theories.
This is indeed the case, though the added term proportional to
$\sqrt{N}$ complicate some parts of the analysis.
The details of the full proof of this statement are presented in appendix \ref{ap:proof}.

We note that the equations \eq{modcon1} and \eq{modcon2}
enable us to identify many uncurable theories with ease.
For example, it can be shown that the $T=0$ theory
with gauge group and matter content
\begin{equation}
SU(7): \;\;\;\;\; \;\;\;\;\;
27 \times {\tiny\yng(1)}+
1 \times {\tiny\yng(2,1,1)},
\;\;\;\;\; (H-V = 351) \,
\label{eq:curable-example}
\end{equation}
is uncurable, since
\begin{align}
H-{V} &> 273-29T + (T^2+6T +8) =281\\
H-{V} &> 273-29T + (2T+4) + \sqrt{2}(T+2)\sqrt{N} = 277+ 2\sqrt{56}= 291.9...
\end{align}

To summarize, we have defined `curable theories' to be supergravity
theories that satisfy the following conditions:
\begin{enumerate}
\item The gauge group is non-abelian.
\item The theory has no singlet hypermultiplets.
\item $H-V >273-29T$
\item The theory can be made non-anomalous by adding $U(1)$
vector fields that are independent degrees of freedom, as well as
possibly adding singlet hypermultiplets.
\item In the resulting non-anomalous theory, there exists a
choice for the scalars in the tensor multiplets
that makes the kinetic terms of all gauge fields positive.
\end{enumerate}

We have proven the following facts:
\begin{enumerate}
\item The number of non-anomalous non-abelian theories is finite \cite{Kumar:2009ae}.
\item The number of $U(1)$'s one can add to non-anomalous theories is finite.
\item The number of curable theories is finite.
\item The number of $U(1)$'s one can add to curable theories is finite.
\end{enumerate}

As pointed out in the beginning of this section, any non-anomalous theory with $U(1)$'s
can be constructed by adding abelian vector multiplets and neutral hypermultiplets to a
non-anomalous or curable theory with no abelian gauge symmetry.
Hence it follows that there is only a finite number of distinct gauge/matter structures
a 6D $(1,0)$ theory could have even when we allow abelian components
to the gauge group.
In particular, this implies that the total rank of the gauge group is bounded,
even when we admit abelian factors in the gauge group.

\subsection{$T \geq 9$} \label{ss:Tgeq9}

In this section, we show that for $T \geq 9$ a bound cannot be imposed
on the number of $U(1)$'s as we have done in the case $T <9$.
We first show that there are certain classes of theories to which
one could add an arbitrary number of $U(1)$'s, and discuss why this is
not possible when $T <9$.
We end with an example of an infinite class of
non-anomalous theories with an unbounded number of $U(1)$'s.

Suppose we have a theory $\sT_0$ with gauge group $\GG_0$ that satisfies all
the anomaly equations and has an $SO(1,T)$ unit vector $j_0$
that satisfies $j_0 \cdot b_\kappa>0$ for all gauge groups $\kappa$.
Denote the number of vector and hypermultiplets of this theory as
$V_0$ and $H_0$.

Suppose an $SO(1,T)$ vector $b$ that satisfies the following conditions exists:
\begin{enumerate}
\item $b$ is light-like, {\it i.e.,} $b^2=0$.
\item $a\cdot b=0$.
\item $b_\kappa \cdot b =0$ for all $\kappa$.
\item $b \cdot j_0 >0$.
\end{enumerate}
Recall that in the case $T<9$ it is impossible for a vector $b$ to satisfy conditions
1, 2 and 4 at the same time. In that case $a$ is a time-like vector
and if 1 and 2 are satisfied, $b$ must be a zero vector. This is what prevented
us from having a $U(1)$ with nothing charged under it.

The situation is quite different when $T\geq 9$; in this case a vector $b$ satisfying the
four conditions above is not ruled out in general.
Once such a $b$ is available one could construct theory $\sT_k$ from $\sT_0$
with the following properties.
\begin{enumerate}
\item The gauge group is $\GG_k = \GG_0 \times U(1)^k$.
\item The matter content is that of $\sT_0$ with $k$ neutral hypermultiplets added.
\item Nothing is charged under the $U(1)$'s, {\it i.e.,} $q_{I,j}=0$ for all $I, j$.
\item The non-abelian anomaly coefficients are given by $b_\kappa$.
\item The abelian anomaly coefficients are given by $b_{ij} = \delta_{ij} b$.
\item The tensor multiplet scalar vacuum expectation value is given by $j_0$.
\end{enumerate}
By adding the $k$ neutral hypermultiplets, the gravitational anomaly condition,
\be
H_k-V_k=(H_0 +k)-(V_0 +k)=H_0 -V_0 =273-29T
\ee
is satisfied. The non-abelian anomaly factorization conditions are all  satisfied
by definition. We find that all the $U(1)$ anomaly equations \eq{one}-\eq{four}
are also satisfied as both sides of the equation turn out to be $0$.
Also,
\be
j_0 \cdot b_{ij} = (j_0 \cdot b) \delta_{ij}
\ee
is a positive definite matrix by the assumption that $b \cdot j_0 >0$.
Therefore, this theory satisfies all the anomaly equations and has a sensible kinetic term.
Since this is true for any $k$ we find that we could add an infinite number of
$U(1)$'s to $\sT_0$.

We conclude this section by presenting an explicit example of
an infinite class of theories with an unbounded number of $U(1)$
factors in the gauge group. The simplest case, when there are no
non-abelian factors, turns out to serve our purpose.
A $U(1)^k$ theory with $273-29T+k$ neutral hypermultiplets and
$a, b_{ij}$ given by
\be
a=(-3,1 \times T,0,\cdots,0),\quad b_{ij} = b \delta_{ij} \quad
\text{for }b=(3,(-1) \times 9,0,\cdots,0)
\ee
satisfies all the factorization equations.
$x \times n$ denotes that $n$ consecutive  components have
the same value $x$.
Defining
\be
j=(1,0,0,0,0,\cdots,0) \,,
\ee
we find that the matrix for the kinetic term of the vector multiplets
\be
j \cdot b_{ij} = 3\, \delta_{ij}
\ee
is positive definite. $k$ is bounded below by $29T-273$ but has no upper-bound.

\section{Infinite Classes of Non-anomalous Theories with $U(1)$'s} \label{s:inf}

In this section, we investigate the second and third  questions
(II and III) posed
in the introduction, beginning with II:
Given the gauge/matter content of the theory---by which we mean
that we fix the gauge group and the representations of the
hypermultiplets with respect to the non-abelian part of the
gauge group---are there an infinite number of solutions
to the $U(1)$ charge equations?
We denote these $U(1)$ charge equations `hypercharge' equations.


As pointed out in the introduction, there are infinite families
of solutions that may be `trivially generated' in the following sense.
There certainly exist solutions of the anomaly equations with gauge group
$\GG = \GG_0 \times U(1)^2$.
In such a case, denoting the charge vectors with respect
to the two $U(1)$'s $\vec{q}_1$ and $\vec{q}_2$,
any linear combination $\vec{Q}=r \vec{q}_1 + s \vec{q}_2$
solves the anomaly equation for $\GG' = \GG_0 \times U(1)$
with the same matter structure.
On top of the anomaly cancellation conditions, we may demand that
additional consistency conditions be obeyed 
\cite{Polchinski,Banks:2010zn,HellermanSharpe,SeibergTaylor}.
Three such conditions are applicable to six-dimensional
supergravity theories with compact $U(1)$ abelian factors:
\begin{enumerate}
\item \textbf{Charge Integrality Constraint :}
All charges of particles should be integral with respect to the minimal
charge of the $U(1)$'s.
\item \textbf{Minimal Charge Constraint :}
The greatest common divisor of the charges of all
particles under each $U(1)$ should coincide
with the minimal charge--or inverse of the periodicity--of the $U(1)$.
\item \textbf{Unimodularity Constraint :}
The string charge lattice spanned by the anomaly coefficients
should be embeddable in a unimodular lattice.
\end{enumerate}
The first and second constraints
do not stop us from generating an infinite familiy because
if the initial theory with $\GG = \GG_0 \times U(1)^2$ satisfied
the charge integrality constraint and the minimal charge constraint,
the new theory would also satisfy this constraint when $r, s$ are
taken to be mutually prime integers.
In many cases the unimodularity constraint does not help either, as we
see shortly.

Let us depict the situation with the simplest example.
For $\GG =U(1)^2$, $T=1$ the following charges on the 246
hypermultiplets of the theory solve the anomaly equations.
Assume that there are 48 hypermultiplets with charge $(0,1)$, 48 hypermultiplets
with charge $(1,0)$, 48 hypermultiplets with charge $(1,1)$ and 102
neutral hypermultiplets.
Written in terms of charge vectors
\be
\vec{q}_1 = (1 \times 96, 0 \times 48, 0\times 102), \qquad
\vec{q}_2 = (0 \times 48, 1 \times 96, 0\times 102)\,,
\ee
where $q \times n$ denotes that $n$ consecutive
components have the same value $q$. The only non-trivial
anomaly equations concerned are
\begin{align}
{1 \ov 6} (48 x_1 ^2 + 48 (x_1 +x_2)^2 + 48 x_2^2) &=
(\alpha_{11} +\ta_{11})x_1^2 + 2 (\alpha_{12} +\ta_{12})x_1 x_2+ (\alpha_{22} +\ta_{22}) x_2^2 \\
{2 \ov 3} (48 x_1^4 + 48 (x_1 +x_2)^4 + 48 x_2^4) &=
(\alpha_{11}x_1^2 + 2 \alpha_{12} x_1 x_2+ \alpha_{22} x_2^2)
(\ta_{11}x_1^2 + 2 \ta_{12} x_1 x_2+ \ta_{22} x_2^2)
\end{align}
Both equations are satisfied by the choice
\be
\alpha_{11}=\alpha_{22}=2\alpha_{12}
=\ta_{11}=\ta_{22}=2\ta_{12}=8 \,.
\ee
Therefore
\be
\vec{Q} = (r \times 48,(r+s)\times 48, s \times 48, 0\times 102)
\ee
satisfy the equations
\begin{align}
{1 \ov 6} (48 r^2 + 48 (r +s)^2 + 48 s^2) &=
16 r^2 + 16 rs + 16 s^2 \\
{2 \ov 3} (48 r^4 + 48 (r +s)^4 + 48 s^4) &=
(8 r^2 + 8rs +8s^2 )^2 \,.
\end{align}
It is easy to see that this choice of charges solves the anomaly equation
for $\GG=U(1)$ with $\alpha =\ta = (8 r^2 + 8rs +8s^2 )$.
Therefore, we obtain an infinite class of solutions to the anomaly
equations for $\GG = U(1)$.

It is clear that imposing the charge integrality constraint and
the minimal charge constraint does not stop us from generating this infinite family
as we may take $r$ and $s$ to be mutually prime integers.
Now we show that the unimodularity constraint is also satisfied
when $r$ and $s$ are integers.

It is useful to notice that when $T=1$,
a sufficient condition for the unimodularity constraint 
is that all the anomaly coefficients
$\alpha$ and $\ta$ defined in section \ref{ss:teq1}
are even integers.
This is because if all $\alpha$ and $\ta$ are even integers,
all string charge vectors
\be
a = \begin{pmatrix} -2 \\ -2 \end{pmatrix}, \quad
b = {1 \ov 2} \begin{pmatrix} \alpha \\ \ta \end{pmatrix}
\ee
are embeddable in the unimodular lattice spanned by
\be
\begin{pmatrix} 1 \\ 0 \end{pmatrix} \quad \text{and} \quad 
\begin{pmatrix} 0 \\ 1 \end{pmatrix}\,,
\ee
with inner product structure $\Omega$ as defined in
(\ref{eq:omega}).
When $r,s$ are integers, $\alpha$ and $\ta$ of the $U(1)$
are both equal to $(8 r^2 + 8rs +8s^2 )$, which is an even integer.
Therefore, the unimodularity constraint does not rule out this
infinite class of theories.

The natural follow-up question to ask is whether there is some
gauge/matter structure that permits an infinite number of distinct
solutions to the hypercharge equations that cannot be lifted to
a theory with more $U(1)$'s.
It turns out that there are infinite classes of solutions to anomaly
equations of a theory with gauge group $\GG_0 \times U(1)$ that
cannot be lifted to $\GG_0 \times U(1)^2$.
The example we examine is the theory with gauge group
$SU(13) \times U(1)$ that we presented at the end of section \ref{ss:teq1}.
There we found a solution to the non-abelian
factorization condition with 4 antisymmetrics, 6 fundamentals and 23
singlets in the $SU(13)$. The non-abelian part of the factorized polynomial is
\be
-{1 \ov 16}(\tr R^2 - 2\tr F_{SU(13)}^2) \wedge
(\tr R^2 - 2\tr F_{SU(13)}^2 ) \,.
\ee
Denoting the charge of hypermultiplets in the
antisymmetric/fundamental/singlet representation as
$a_x$($x=1,\cdots,4$)/$f_y$($y=1,\cdots,6$)/$s_z$($z=1,\cdots,23$)
the anomaly equations become
\begin{align}
9 \sum_x a_x + \sum_y f_y &= 0 \label{su131}\\
78 \sum_x a_x^4 + 13 \sum_y f_y^4 + \sum_z s_z^4 &={3 \ov 2} \alpha \ta \label{su132}\\
78 \sum_x a_x^2 + 13 \sum_y f_y^2 + \sum_z s_z^2 &=6\alpha + 6 \ta \label{su133}\\
44 \sum_x a_x^2 + 4 \sum_y f_y^2 &= 2 \alpha + 2 \ta \label{su134}
\end{align}

There is an ansatz that solves this equation given by
\begin{align}
(a_x) &= (-3a -{2 \ov 3}f,a,a,a) \\
(f_y) &=(f,f,f,f,f,f) \\
(s_z) &= ((6a+f)\times 18,0,0,0,0,0)
\end{align}
where in the last line we mean that 18 of the $s_z$ take the
value $(6a+f)$ while five take 0. This ansatz satisfies the first
equation and renders the third and fourth equations equivalent.
Then the second and third equation can be solved with respect to
$\alpha, \ta$ to yield
\be
{\alpha \ov f^2}, {\ta \ov f^2} = {2 \ov 9}
\left[ (49 + 198t+594 t^2) \pm \sqrt{39} (1+6t) \sqrt{23-24t-36t^2} \right]
\label{ata}
\ee
where we have defined $t=a/f$.
It is easy to see that $\alpha, \ta$ are real as long as
\be
{-2 - 3\sqrt{3} \ov 6} \leq t \leq {-2 + 3\sqrt{3} \ov 6} \,.
\ee
Both $\alpha$, $\ta$ are positive when $t$ is in this range.
Hence we see that there are an infinite number of integral hypercharge
solutions to the equations (\ref{su131})-(\ref{su134}) that give allowed
values of $\alpha, \ta$.

It is clear that this theory cannot be lifted to a theory with
gauge group $SU(13) \times U(1)^2$. Although the ansatz for the
given solution seems to imply that this theory can be lifted, for
example by choosing the charges for
one $U(1)$ to be proportional to $a$ and the charges for the other
$U(1)$ to be proportional to $f$, the fact that
$a/f$ must lie in a certain range implies that there must
be an obstruction to doing this. The obstruction is that if one tries to
lift the theory to a theory with gauge group $SU(13) \times U(1)^2$,
the matrices $\alpha_{ij}$ and $\ta_{ij}$ of this theory
cannot be made into positive definite real matrices
as is required for the $U(1)$ gauge fields to have
positive-definite kinetic terms.

The next question to ask is whether there is an infinite subclass of
these theories that satisfy all three quantum consistency conditions
introduced at the beginning of this section.
Generating a subclass of theories that satisfy the integrality constraint
and the minimum charge constraint is not difficult.
For example, by taking $f$ and $a$ to be mutually prime integers
and $f$ to be a multiple of 3, one can generate an infinite class of solutions that
satisfy these two constraints.
These conditions, however, do not lead to the unimodularity constraint.

In order to construct a subclass of theories that satisfy all three
constraints, let us examine whether there exists an infinite
number of rational values of $t$ that make the right hand side of (\ref{ata})
rational. This problem boils down to the question of whether the equation
\be
23-24t-36t^2 = 39 q^2
\ee
admits an infinite number of solutions with rational $t$ and $q$.
We find that there indeed are an infinite number
of rational solutions to this equation
using methods outlined in chapter 7 of \cite{Mordell}.
When
\be
{a \ov f}=t = {13k^2 -234 k -51 \ov 24(13k^2 +3)}
\ee
for  $k$ rational we find that
\begin{align}
{\alpha \ov f^2} &= {13 \ov 144(3+13k^2)^2}
\left[ 6687 + 54756 k + 94458 k^2 -124956 k^3 + 39455 k^4 \right]\\
{\ta \ov f^2} &= {13 \ov 144(3+13k^2)^2}
\left[ 2475 + 37908 k + 170274 k^2 -29484 k^3 + 9035 k^4 \right] \,.
\end{align}
Hence we find that the number of non-anomalous
theories with $SU(13) \times U(1)$ with this particular type
of matter content is infinite.

To be clear, we now spell out the explicit subclass of theories that satisfy
all three quantum consistency conditions.
Setting $k=r/s$ for integers $r$ and $s$ in the above equations, we find that when
\begin{align}
a&=13r^2-234rs-51s^2 \\
f&=24(13r^2+3s^2) \,,
\end{align}
$\alpha$ and $\ta$ take on the values
\begin{align}
\alpha  &= {52}
\left[ 6687s^4 + 54756 s^3 r + 94458 s^2r^2 -124956 sr^3 + 39455 r^4 \right]\\
\ta &= {52}
\left[ 2475 s^4 + 37908 s^3r + 170274 s^2r^2 -29484 sr^3 + 9035 r^4 \right]\,,
\end{align}
which are even integers.
As discussed early on in this section, this implies that the
string charge lattice can be embedded in a unimodular lattice.
It is clear that this ansatz assigns integer charges to all the fields
and hence the charge integrality constraint is also satisfied.
If $a$ and $(-3a-2f/3)$ are mutually prime, the minimal charge constraint is also
satisfied. There are an infinite number of integer pairs $(r,s)$ that render
$a$ and $(-3a-2f/3)$ mutually prime.
In fact, we can show that when
\begin{align}
r&=84 n+43 \\
s&=182n+92
\end{align}
for integer $n$, $a$ and $(-3a-2f/3)$ are mutually prime.
This fact is proven in appendix \ref{ap:prime}.

We have found a particular gauge/matter structure with
one $U(1)$ that has an infinite number of distinct solutions
to the hypercharge equations for $T=1$. Furthermore the theory
cannot be lifted to a theory with two $U(1)$'s for
these hypercharge assignments.

The situation is rather subtle for the case of $T=0$. The equations
\eq{pi10}-\eq{pi40} make it clear that any infinite class of solutions
to the anomaly equation with charge vectors of the form
\be
\vec{Q} = r\vec{q}_1 + s \vec{q}_2
\ee
for one $U(1)$ can be lifted to $U(1)^2$. As in the $T=1$ case there
are a plethora of examples of gauge/matter structure that admit an
infinite family of hypercharge solutions in this way.
If, however, we want to identify an infinite class of theories that satisfy anomaly
equations for a single $U(1)$ factor that cannot be extended to $U(1)^2$,
we cannot have a simple linear ansatz as in the $T=1$ case.
Examining some specific examples of $T = 0$
theories gives interesting number theory
problems that in some cases seem to have infinite $U(1)$ families that cannot
be extended to $U(1)^2$ models, but we do not go into the details of
these constructions here.


\section{Conclusions} \label{s:conclusion}

We have considered 6D supergravity theories with $(1,0)$ supersymmetry
with abelian as well as nonabelian gauge group factors. The following
statements have been proven for such theories when the number of
tensor multiplets $T$ satisfies $T < 9$:
\begin{enumerate}
\item The number of abelian vector multiplets is bounded above
by \eq{centralprime0} and \eq{centralprime}. The upper bound
is determined by the nonabelian gauge/matter content.
\item The number of possible gauge groups and nonabelian matter
content is finite, though there are families with infinite numbers of
possible distinct $U(1)$ charges.
\end{enumerate}
From (2), it immediately follows that
\begin{enumerate}
\setcounter{enumi}{2}
\item There is a global bound on the rank of the gauge group
of any non-anomalous 6D ${\cal N} =  (1,0)$ theory with $T < 9$.
\end{enumerate}

The conclusions we have reached for theories with abelian factors are
in some ways closely parallel to the analogous results bounding the
space of 6D supergravity theories with only nonabelian factors
 \cite{Kumar:2009ae, Kumar:2009ac}. Adding abelian factors does not
change the basic result that the set of possible gauge groups and
nonabelian matter representations is finite for $T < 9$. The biggest
difference for theories with abelian gauge group factors is that we
cannot bound the number of distinct possibilities for $U(1)$ charges.
Even this result, however, fits naturally into the pattern of theories
with nonabelian gauge groups for small $T$. In \cite{Kumar:2010am},
we analyzed the set of allowed 6D theories with $SU(N)$ gauge groups
and no tensor fields ($T = 0$). For large $N$ the bounds on the set
of allowed representations of $SU(N)$ under which matter fields
transform are quite stringent. As $N$ decreases, however, more and
more exotic matter representations are allowed by the anomaly
conditions and other known low-energy constraints. For $SU(3)$ there
are over 10,000 different matter combinations possible, and for
$SU(2)$ there are many millions of combinations possible, including,
for example,
matter in the 113-index symmetric tensor representation ({\bf 114}).
The infinite range of possible $U(1)$ charges seems like a natural
divergent limit to this range of theories.

These results naturally lead to the question of whether it is possible
to place stronger bounds on the set of consistent 6D theories than
those understood from anomaly cancellation and other known constraints,
either from string theory or from other macroscopic considerations.
For theories without abelian factors, this question has led to an
improved understanding of the space of 6D theories. F-theory
 \cite{MV, Vafa-F-theory} gives a method for constructing a very
general class of 6D theories. By relating discrete structure of the
low-energy theory to topological structure of F-theory constructions,
additional constraints placed by F-theory on the low-energy 6D theory
have been identified for theories with nonabelian gauge groups
 \cite{Kumar:2009ac}. One such condition is that the dyonic string
charge lattice of the low-energy theory, which contains the anomaly
lattice spanned by the $SO(1, T)$ vectors $b_i$, must be self-dual.
This condition was shown to be a requirement for quantum consistency
of any 6D supergravity theory in \cite{SeibergTaylor}. A second
condition that is imposed by F-theory on supersymmetric theories is
the Kodaira constraint that the total elliptically fibered space be
Calabi-Yau. For theories with $T = 0$ and $SU(N)$ gauge group, for
example, this constraint implies that $-12a =36 \geq N b$, where $b$
is the anomaly coefficient for the $SU(N)$ gauge group. This
condition places an additional constraint on the set of possible
$SU(N)$ theories beyond the conditions imposed by anomaly
cancellation. While all $T = 0$ $SU(N)$
models with $N > 8$ that satisfy
anomaly cancellation also automatically satisfy this Kodaira
constraint, this constraint places increasingly strong additional
restrictions on $SU(N)$ theories for small $N$. In particular, this
condition reduces the millions of possible models with exotic matter
representations for theories with $SU(2)$ gauge group to less than 200
models, with the largest representation appearing being the ${\bf 6}$
of $SU(2)$ \cite{Kumar:2010am}. More generally, the Kodaira
constraint rules out all known infinite families of 6D theories with
nonabelian gauge groups, even for $T \geq 9$, consistent with the
known fact that there are a finite number of different possible gauge
groups and matter content for F-theory constructions
 \cite{Kumar:2009ac}. It is an open question whether (a) the Kodaira
constraint can be realized from the point of view of 6D supergravity
as a general consistency condition on any quantum theory,  (b) this
constraint depends crucially on the UV-completion and represents a
constraint intrinsic to string theory, or (c)  there are other more
exotic string constructions beyond F-theory that can realize theories
violating the Kodaira constraint.

Since F-theory can also only allow a finite number of possible $U(1)$
charges for any class of theories, it seems that a constraint
analogous to the Kodaira constraint must hold for theories with
abelian gauge group factors. Such a constraint should place bounds on
the $U(1)$ anomaly coefficients $b_{ij}$, leading to constraints on
abelian charges through equation\ (\ref{one}). The origin of such a
constraint in F-theory is less clear for abelian factors than for
nonabelian factors, however, since abelian factors arise in a global
and less transparent fashion in F-theory than nonabelian factors. 
Some  progress in formulating abelian factors geometrically in
F-theory that may be relevant to this problem will appear in \cite{Daniel-intersection}.

Another question whose resolution may help shed light on the issues
addressed by this paper is the determination of the precise limit on
the number of $U(1)$ factors that may arise in a 6D supergravity
theory with fixed nonabelian gauge group and matter structure.  The
simplest case of this question is when $T = 0$ and there is no
nonabelian gauge group.  We know that F-theory models exist with up to
$k = 7$ $U(1)$ factors, though the explicit geometry is only known up
to $k = 3$, and the upper bound found here of $k < 17$ is probably not
optimal.  It would be interesting to find methods for decreasing the
upper bound and/or constructing explicit F-theory models with larger
$k$ both in this simplest case and more generally.
Recently, an impressive list of toric Calabi-Yau threefolds
that are elliptic fibrations over $\field{P}^2$ have been put together
in \cite{Braun}. Obtaining an upper bound on $k$ for
F-theory compactifications on these manifolds
seems to be a goal attainable in the near future.

Finally, understanding $U(1)$ factors in supergravity and string
constructions presents a similar challenge in four dimensions, though
with additional subtleties.   It seems likely that further progress on
understanding $U(1)$ factors in six dimensions will also shed light on
the 4D problem.

\section*{Acknowledgements}

We would like to thank Allan Adams, Mboyo Esole, I\~naki
Garc\'ia-Etxebarria, Thomas Grimm,
James Halverson, Vijay Kumar, Joe Marsano,
John McGreevy,
and David Morrison for helpful discussions.
DP would especially like to thank Koushik Balasubramanian for
kindly listening to and commenting on his ramblings on the subject of
this paper. We would like to thank the Ohio State University physics department
and the organizers of the 2010 String Vacuum Project Fall Meeting for
hospitality during the course of writing this paper.
DP acknowledges support as a String Vacuum
Project Graduate Fellow, funded through NSF grant PHY/0917807.  This
research was supported by the DOE under contract \#DE-FC02-94ER40818.

\appendix

\section{Proof of Bound on Curable Theories} \label{ap:proof}

We  prove that the number of curable theories as
defined in section \ref{ss:curability} is finite for $T < 9$.
The crucial fact we use is that for curable theories
one of the two following conditions must hold:
\begin{align}
H-{V} &\leq 273-29T + (T^2+6T +8) \label{modconA1} \\
H-{V}-\sqrt{2}(T+2) \sqrt{N} &\leq 273-29T + (2T+4) \label{modconA2}
\end{align}
where $N$ is the number of hypermultiplet representations of the theory.
It is clear from \cite{Kumar:2009ae,Kumar:2010ru} that there could not
be an infinite family of theories for which the first condition holds
as it requires $H-V$ to be bounded. Therefore, it is sufficient
to show that there does not exist an infinite family of curable theories
for which the second condition \eq{modconA2} holds.
It proves convenient to define
\be
c(T) \equiv {T+2 \ov \sqrt{2}} \,.
\ee

Before presenting the proof of the desired result,
we will point out that the proof is very similar to that given for
non-abelian theories in \cite{Kumar:2009ae,Kumar:2010ru}.
Proofs of the existence of bounds on non-anomalous theories
are carried out by two steps in these references.
First, the authors identify infinite classes of theories
that satisfy all the anomaly equations other than the gravitational
anomaly bound, and that have
positive kinetic terms for the gauge fields.
Then they show that it is impossible for all the theories
in that infinite class to satisfy the gravitational anomaly bound.
This is proven by showing that as the total rank of the gauge
group increases, the increase of $H$ is much faster than $V$.
This proves that constructing an infinite class of theories
that satisfy all the anomaly cancellation conditions and that have
positive kinetic terms for the gauge fields does not exist.

We also take the same approach in proving our bounds.
In our case, however, we must prove
that the increase of $H-c(T)\sqrt{N}$
is much faster than $V$ for the infinite classes of theories
one could construct. Most of our effort will be put
in to showing that $\sqrt{N}$ does not increase so fast as
to affect the growth of $H$.

There are infinite classes of theories that this is easy to show.
For example, for the class of theories whose $H$ and $V$
exhibit a scaling behavior with respect to the rank of
the total gauge group when it becomes large,
the arguments presented in  \cite{Kumar:2009ae,Kumar:2010ru}
can be virtually repeated.
This is because equation \eq{modconA2} implies that
\be
\label{modconA2prime}
(\sqrt{H}-c(T))^2-V \leq 277 -27T+c(T)^2 \,.
\ee
This is because as we have assumed there exist
no singlets in curable theories, and hence
\be
N \leq {H \ov 2} < H
\ee
holds.
Therefore, the scaling behavior of
$(\sqrt{H}-c(T))^2-V$ and $H-V$ with respect to the rank is equivalent
and the boundedness argument for these classes of theories
are essentially the same.
In particular, for an infinite class of theories whose simple group
factors have bounded rank, the proof of boundedness
given in \cite{Kumar:2009ae,Kumar:2010ru} can be used
with very little adjustments.
This is presented in section \ref{ss:case1}.

It is, however, worth pointing out that for some infinite class of theories,
the situation is rather subtle.
When there exist simple group factors with unbounded rank
in the infinite class of theories,
the bound \eq{modconA2prime}
becomes too delicate to use.
In that case the stronger bound \eq{modconA2} turns out
to be more useful in proving the existence of bounds of
curable theories.
We will carry this out in section \ref{ss:case2}.

We now turn to presenting the complete proof of the
bound on curable theories.
We proceed by {\it reductio ad absurdum}.
Let us assume there is an infinite family of curable non-abelian theories
with gauge group $\prod_\kappa \GG_\kappa$.
Due to the bound on ${H}-{V}-2c\sqrt{N}>(\sqrt{H}-c)^2-V-c^2 $, we see that theories in an
infinite family of curable theories should be unbounded in the dimension of the gauge group.
If not, $H$ and $V$ are both bounded,
and hence only a finite number of theories can be constructed.
There are two ways  an unbounded family can occur. These are given as the following:
\begin{enumerate}
\item The dimension of each $\GG_\kappa$, or equivalently,
$d_{\text{Adj}_\kappa}$ is bounded,
but the number of simple factors is unbounded in this family.
\item The dimension of a single simple group factor $\GG_\kappa$ is unbounded.
\end{enumerate}
We show that both kinds of families cannot exist in the following subsections.

An important fact we use throughout the proof is the fact that,
\begin{align}
\label{nhlb}
\begin{split}
H 
&\geq
\begin{pmatrix}
\text{Number of pairs of $\GG_\iota \neq \GG_\kappa$} \\
\text{for which there exists a jointly charged hypermultiplet.}
\end{pmatrix}
\\
&= (\text{Number of pairs of $\GG_\iota \neq \GG_\kappa$ with $b_\iota \cdot b_\kappa \neq 0$.})
\end{split}
\end{align}
This can be shown in the following way.

Suppose a hypermultiplet representation $I$ is charged under $\lambda \geq 2$
gauge groups. Then
\be
\MM_I \geq 2^\lambda \geq \lambda(\lambda-1) =
\begin{pmatrix}
\text{Number of pairs of $\GG_\iota \neq \GG_\kappa$} \\
\text{that $I$ is charged jointly under.}
\end{pmatrix}
\ee
and therefore
\begin{align}
\begin{split}
H = \sum_I \MM_I
&\geq \sum_I
\begin{pmatrix}
\text{Number of pairs of $\GG_\iota \neq \GG_\kappa$} \\
\text{that $I$ is charged jointly under.}
\end{pmatrix} \\
&\geq
\begin{pmatrix}
\text{Number of pairs of $\GG_\iota \neq \GG_\kappa$} \\
\text{for which there exist a jointly charged hypermultiplet.}
\end{pmatrix}
\end{split}
\end{align}
This means that any ordered pair of gauge groups that has matter jointly charged
under it contributes at least 1 to $H$. This proves the inequality in the
first line of \eq{nhlb}.

Since $A^R_\kappa$ for any representation $R$ of any simple Lie group $\GG_\kappa$
is positive and since,
\be
b_\iota \cdot b_\kappa =
\sum \lambda_\kappa \lambda_\iota \sum_I \MM_I^{\iota \kappa} A^I_\iota A^I_\kappa \geq 0
\ee
the necessary sufficient condition for two gauge groups $\GG_\iota, \GG_\kappa$
to have jointly charged matter is $b_\iota \cdot b_\kappa \neq 0$.
Therefore
\begin{align}
\begin{split}
H \geq \sum_{\iota \neq \kappa,~ b_\iota \cdot b_\kappa \neq 0} 1
=(\text{Number of pairs of $\iota \neq \kappa$ with $b_\iota \cdot b_\kappa \neq 0$.})
\end{split}
\end{align}
This proves the equality in the second line of \eq{nhlb}.

\subsection{Case 1 : Bounded Simple Group Factors} \label{ss:case1}

Let us assume that there exists an infinite number of curable theories
with bounded simple group factors but with unbounded total dimension.

Let's denote the gauge group of this infinite family of theories $\{ \sT_\Phi \}$ as
$\GG_\Phi = \prod_{\kappa=1}^\nu \GG_\kappa$ with $d_{\text{Adj}_\kappa} < D$.
Notice that we are denoting the number of gauge group factors, $\nu$.
So as $\Phi \rightarrow \infty$, $\nu \rightarrow \infty$.
It is useful to classify the gauge group factors into three types
according to their $b^2$ value:
\begin{enumerate}
\item \textbf{Type Z : }$b_\kappa^2=0$
\item \textbf{Type N : }$b_\kappa^2<0$
\item \textbf{Type P : }$b_\kappa^2>0$
\end{enumerate}

Since the dimension of each factor is bounded,
\be
(\sqrt{H}-c)^2 \leq H-2c\sqrt{N}+c^2 \leq 273-29T+\nu D+c^2 \equiv B \sim \OO(\nu) \,.
\ee
Therefore, the dimension of any representation is bounded also by
\be
B' \equiv (\sqrt{B}+c)^2 \sim \OO(\nu) \,.
\ee


Let's denote the number of $N, Z, P$ type factors as $\nu_N, \nu_Z, \nu_P$.
Then
\be
\nu= \nu_N + \nu_Z + \nu_P \,.
\ee
It is shown in \cite{Kumar:2010ru} that $b_\iota, b_\kappa$ of
any two P type factors $\GG_\iota, \GG_\kappa$ satisfy
$b_\iota \cdot b_\kappa > 0$. Also it is shown that there exists
$\nu_N^2/T-\nu_N$ distinct ordered pairs of type N gauge group factors
that have matter jointly charged under them. Therefore, using \eq{nhlb}
we can show that
\be
\nu_P (\nu_P-1) + ({\nu_N^2 \ov T} -\nu_N)\leq H \leq B' \sim \OO(\nu) \,.
\ee
Thus when $\nu$ is large,
\be
\nu_P, \nu_N \leq \OO(\sqrt{\nu}) \ll \nu
\ee
and therefore the majority of gauge group factors are
of type $Z$:
\be
\label{ordz}
\nu_Z \sim \OO(\nu)
\ee

From the fact that two lightlike vectors cannot have zero
inner product unless they are parallel, it is clear that in order for
two Z type gauge groups to
have no jointly charged matter their $b$ vectors
must be parallel.
When we denote the size of the largest collection of parallel
type Z vectors as $\mu$ there are at least $\nu_Z (\nu_Z-\mu)$
ordered pairs of type Z gauge groups with $b_\iota \cdot b_\kappa \neq 0$.
This means that
\be
\nu_Z (\nu_Z-\mu) = (\nu_Z-\mu)(\nu-\nu_P-\nu_N) \leq B' \,,
\ee
so from \eq{ordz} we see that $\nu_Z-\mu$ is of order at most $\OO(1)$,
{\it i.e.,} $\nu_Z-\mu$ is bounded
as a function of $D$ as we take $\nu \rightarrow \infty$.
Therfore
\be
\mu \sim \nu_Z \sim \OO(\nu) \,.
\ee

Meanwhile, it was shown in \cite{Kumar:2009ae} that all Z factors satisfy $H -V>0$
on their own. Since the $\mu$ Z factors have no jointly charged matter among themselves,
\be
H-V >
\mu-D \times [(\nu_Z-\mu)+\nu_p+\nu_N] \sim \OO(\nu) \,,
\ee
{\it i.e.,} the right hand side of the inequality is unbounded as a function of $\nu$.
Then it is clear that,
\be
H-V-2c\sqrt{N} > H-2c\sqrt{H}-V=(\sqrt{H}-c)^2-V-c^2
\ee
is also unbounded as a function of $\nu$.
Therefore, $H-V-2c\sqrt{N}$ cannot be bounded when
each simple group factor of the infinite family has bounded rank.
This rules out case 1.


\subsection{Case 2 : Unbounded Simple Group Factors} \label{ss:case2}

Let us assume that there exists an infinite number of curable theories
with a simple group factor that is unbounded.
This is possible if the gauge group contains a classical group $H(\NN)$
(which is either $SU$, $SO$ or $Sp$) with unbounded rank.
In this case, there would be an infinite subfamily whose gauge group is given by
$H(\NN)\times \GG_\NN$ with fixed classical group type $H$, and
$\GG_\NN=\prod_{\kappa=1}^{\nu(\NN)} \GG_\kappa$ an
arbitrary product of simple gauge groups with $\NN$ unbounded.
It is shown in \cite{Kumar:2009ae}
that when $\NN$ is large, the $\HH(\NN)$ block must be
among those given in table \ref{t:matterN}.

\begin{table}[t!]
\centering
\setlength{\extrarowheight}{3pt}
 \begin{tabular}{|c|c|c|c|c|c|} 
 \hline
 Group & Matter content & $H'-V'$ & $2c\sqrt{N'}$ & $a\cdot b$ & $b^2$ \\ \hline
 \multirow{4}{*}{$SU(\NN)$} & $2\NN\ {\tiny\yng(1)}$ & $\NN^2+1$& $\leq 2c\sqrt{2\NN}$ & 0 & -2\\
 & $(\NN+8)\ {\tiny\yng(1)}+1\ {\tiny\yng(1,1)}$ & $ \frac{1}{2}\NN^2+\frac{15}{2}\NN+1 $& $\leq 2c\sqrt{\NN+9}$ & 1 &-1\\
 & $(\NN-8)\ {\tiny\yng(1)}+1\ {\tiny\yng(2)}$ & $ \frac{1}{2}\NN^2-\frac{15}{2}\NN+1 $& $\leq 2c\sqrt{\NN-7}$ & -1 & -1\\
 & $16\ {\tiny\yng(1)}+2\ {\tiny\yng(1,1)}$ & $ 15\NN+1 $ & $\leq 2c\sqrt{18}$ & 2 & 0\\ \hline
 $SO(\NN)$ & $(\NN-8)\ {\tiny\yng(1)}$ & $\frac{1}{2}\NN^2-\frac{15}{2}\NN$ & $\leq 2c\sqrt{\NN-8}$ & -1 & -1\\ \hline
 \multirow{2}{*}{$Sp(\NN/2)$} & $(\NN+8)\ {\tiny\yng(1)}$ & $\frac{1}{2}\NN^2+\frac{15}{2}\NN$ & $\leq 2c\sqrt{\NN+8}$ & 1 & -1\\
 & $16\ {\tiny\yng(1)}+1\ {\tiny\yng(1,1)}$ & $15\NN$ & $\leq 2c\sqrt{17}$ & 2 & 0\\ \hline
 \end{tabular}
 \caption{Allowed charged matter for an infinite family of models with gauge group $H(\NN)$.
 The last column gives the values of $\alpha,\ta$ in the factorized anomaly polynomial. $H'$, $V'$ and $N'$
 are as defined in the text.}
\label{t:matterN}
\end{table}

Let us enumerate the hypermultiplet representations charged under $H(\NN)$
with indices, $I'=1,\cdots,N'$ and the ones uncharged(and hence charged only under other
gauge group factors) as $I''=N'+1,\cdots,(N'+N'')$. Note that $(N'+N'')=N$.
We call the former hypermultiplets $I'$ hypermultiplets and the
latter $I''$ hypermultiplets. We also define
\begin{align}
H' &\equiv \sum_{I'} \MM_{I'}  & V' &\equiv d_{\text{Adj}_{H(\NN)}} \\
H'' &\equiv \sum_{I''} \MM_{I''} & V'' &\equiv \sum_\kappa d_{\text{Adj}_{\GG_\kappa}}
\end{align}
Then
\begin{align}
H-V &= (H'-V')+(H''-V'') \\
H\!-\!V\!-2c\sqrt{N} &\geq
(H'\!-\!V'\!-2c\sqrt{N'} )+(H''\!-\!V''\!-2c\sqrt{2N''} ) \,,
\end{align}
where we have used the fact that for positive $x$ and $y$, $\sqrt{x+y} < \sqrt{x} +\sqrt{y}$.

From the $H-V-2c\sqrt{N}$ constraint,
we see that additional gauge groups have to be added
to make the theory curable since for large $\NN$,
$(H'-V'-2c\sqrt{N'}) \sim \OO(\NN^2)$
or $\sim \OO(\NN)$.
We shortly see that this is not possible for arbitrary large $\NN$.
We explicitly work out the proof for the cases when
$(H'-V'-2c\sqrt{N'}) \sim \OO(\NN^2)$,
but the proof generalizes to the other case straightforwardly.

Let us see whether we can find an infinite class of $\GG_\NN$
such that $(H''-V''-2c\sqrt{N''}) \sim - \OO(\NN^2)$ for
large $\NN$. 
There are again two kinds of behavior of
$\GG_\NN$ under $\NN \rightarrow \infty$.
It can consist of simple gauge factors of bounded rank,
or it can have a simple gauge factor whose rank is unbounded.
We consider these cases separately.

\subsubsection{Case 2-1 : Rank of Simple Gauge Group Factors of $\GG_\NN$ Bounded}

Assume that the dimensions of of the simple gauge group
factors are bounded by $D$. Denoting $\nu(\NN)$ as the number
of gauge group factors, as previously mentioned,
we see that
\be
H''-V'' -2c\sqrt{N} > (\sqrt{H''}-c)^2-c^2 -D\nu(\NN)
\ee
must behave as $-\OO(\NN^2)$ for large $\NN$.
Therefore,
\be
\nu(\NN) \geq \OO(\NN^2) \,.
\ee
Hence we must have an infinite family of theories where the number
of simple gauge group factors of $\GG_\NN$ increase at least as $\OO(\NN^2)$.
Also it is clear that
\be
H'' \leq \OO(\nu)
\ee
for large $\NN$ and therefore,
\be
\label{nhppbd}
H = H'+H'' \leq \OO(\NN^2) + \OO(\nu) \leq \OO(\nu) \,.
\ee

Meanwhile, we know from table \ref{t:matterN}
that the size of the representation
of $I'$ hypermultiplets with respect to the $\kappa$ gauge groups
can be at most of order $\OO(\NN)$ in our case.\footnote
{When $(H'-V'-2c\sqrt{N'}) \sim \OO(\NN)$, the number
of $I'$ hypermultiplets can be at most of order $\OO(1)$.}
Therefore, the maximum number of gauge groups an $I'$ hypermultiplet
could be charged under is given by $\OO(\log \NN) \leq \OO(\log \nu)$.
Since there are at most $\OO(\NN \log \NN) \leq \OO(\sqrt{\nu} \log \nu)$ 
such factors, there exists an order $\OO(\nu) \geq \OO(\NN^2)$
number of gauge groups
among the $\nu$ gauge groups $\GG_\kappa$ under
which only the $I''$ hypermultiplets are charged.
We denote theses gauge groups as $\{ \GG_{\kappa''} \}$.

Let us denote the size of the set $\nu''$
and the total number of vector multiplets
in $\{ \GG_{\kappa''} \}$ as $V'''$.
Note that $\nu'' \sim \OO(\nu)$.
Then
\begin{align}
\label{prep}
\begin{split}
H''-V'' -2c \sqrt{N''}
\geq H''-V'''-2c\sqrt{N''}-D\OO(\sqrt{\nu} \log \nu) \,.
\end{split}
\end{align}

Defining the number of P, N, and Z type factors in $\{ \GG_{\kappa''} \}$
as $\nu_P'', \nu_N''$ and $\nu_Z''$ as before repeating the steps
of case 1 we can show that
\be
\nu_P'', \nu_N'' \leq \OO(\sqrt{\nu''}) \ll \nu'' \,,
\ee
and therefore that
\be
\nu_Z'' \sim \OO(\nu'') \,.
\ee
Also, denoting the size of the largest collection of parallel
type Z vectors as $\mu''$ we may show that
\be
\mu'' \sim \OO(\nu'')
\ee
as in case 1.
We may finally show as in case 1 that
\begin{align}
\begin{split}
H''-V'''
& >\mu''-D \times [(\nu_Z-\mu'')+\nu_P''+\nu_N''] \\
&\sim \OO(\nu'') \sim \OO(\nu) \,,
\end{split}
\end{align}
and hence that
\be
H''-V'''-2c\sqrt{N''} > (\sqrt{H''}-c)^2-V'''-c^2 \geq \OO(\nu) \,.
\ee
Putting this result together with \eq{prep} we find that
\be
H''-V'' -2c\sqrt{N''} > \OO(\nu)-D\OO(\sqrt{\nu} \log \nu) \sim \OO(\nu) \,.
\ee
Hence $(H''-V'' -2c\sqrt{N''}) $ cannot behave as
$-\OO(\NN^2)$ for large $\NN$.
We have come a long way to show that there exists a simple gauge group factor
in $\GG_\NN$ that is unbounded in rank.

\subsubsection{Case 2-2 : A Simple Gauge Group Factor of Unbounded Rank in $\GG_\NN$}

In this case, there must be an infinite family of theories with
\be
\hat{H}(\NN) \times H(\PP) \times \GG_{\NN,\PP}
\ee
with unbounded $\NN$ and $\PP$
where $\hat{H}$ and $H$ are given classical groups.
It is clear that both gauge groups have to come from
table \ref{t:matterN}.

Unless $H'-V'-2c\sqrt{N'}$ for $\hat{H}(\NN) \times H(\PP)$ is bounded,
by the same arguments as case 2-1 we can show that $\GG_{\NN,\PP}$
contains a gauge group factor of unbounded rank.
By the same investigation as in \cite{Kumar:2010ru} we find that
all combinations that have bounded $H'-V'-2c\sqrt{N'}$ cannot
have positive definite kinetic terms.

Hence we are led to the conclusion that there must be an infinite family
of theories with
\be
\tilde{H}(\NN) \times \hat{H}(\PP) \times {H} (\mathcal{Q}) \times \GG_{\NN,\PP,\mathcal{Q}} \,.
\ee
$\NN$, $\PP$ and $\mathcal{Q}$ are unbounded
and $\tilde{H}$, $\hat{H}$ and $H$ are given classical groups.
All three unbounded gauge groups must be from table \ref{t:matterN}.

It also is the case that $H'-V'-2c\sqrt{N'}$ for
$\tilde{H}(\NN) \times \hat{H}(\PP) \times {H} (\mathcal{Q}) $ must be bounded
in order for $\GG_{\NN,\PP,\mathcal{Q}}$ to have no gauge group
factor of unbounded rank. It turns out to be impossible to find
such a family with bounded $H'-V'-2c\sqrt{N'}$ for which there
exists a $j$ vector that gives a positive kinetic term.

Our proof is concluded by the fact that one cannot construct an infinite
family of theories that consists of four blocks from table \ref{t:matterN}
for which all the ranks of the individual gauge group factors go to infinity. $\Box$

\section{A Bound on the Number of Vector Multiplets for \\ Pure Abelian Theories with $T=0$} \label{ap:purebound}

In the case that $T=0$ and the gauge group is purely abelian,
we can obtain a lower bound on
the number of charged hypermultiplets as
a function of the the total rank of the gauge group.
Similar bounds may be obtained for other values of $T<9$ though they are
less stringent.

We label the $U(1)$ gauge groups by $i=1,\cdots,V_A$.
The gravitational anomaly condition imposes that the number of
hypermultiplets is equal to $V_A+273$.
We denote the number of charged hypermultiplets to be $X \leq (V_A+273)$,
and label them by $I=1,\cdots,X$.

For  $T=0$, the vectors
\be
\vec{q}_i \equiv (q_{1,i}, q_{2,i}, \cdots, q_{X,i}) \,,
\ee
whose components are the charges
of the $X$ charged hypermultiplets
under $U(1)_i$, must satisfy
\be
108 \sum_{I} f_I(x_i)^4 =( \sum_I f_I(x_i)^2)^2 \,. \label{e}
\ee
This follows from \eq{pi20} and \eq{pi40}
where, as before, we have defined
\be
f_I(x_i) = q_{I,i} x_i \, .
\ee

In order for the kinetic term matrix proportional to
\be
b_{ij} = \sum_I q_{I,i} q_{I,j}
\ee
to be positive definite,
$\vec{q}_i$ must be linearly independent.
This was explained at the end of section \ref{ss:factcond}.
Therefore, using the $GL(V_A)$ invariance of the equation,
we can redefine $\vec{q}_i$ so these vectors become orthogonal.
It is convenient to normalize them to have norm $\sqrt{108}$, {\it i.e.,}
\be
\sum_I q_{I,i} q_{I,j} = \sqrt{108} \delta_{ij} \,.
\label{qnorm}
\ee
(Note that the $q's$ are not necessarily integers in this basis.)

Plugging this into \eq{e} and expanding,
we find that $q_{I,i}$ must also satisfy
\begin{align}
\sum_I q_{I,i}^4 &= 1 \\
\sum_I q_{I,i}^2 q_{I,j}^2 &= {1 \ov 3} & &\text{for $i,j$ distinct} \\
\sum_I q_{I,i}^3 q_{I,j} &=0 & &\text{for $i,j$ distinct} \\
\sum_I q_{I,i}^2 q_{I,j} q_{I,k} &=0 & &\text{for $i,j,k$ distinct} \\
\sum_I q_{I,i} q_{I,j} q_{I,k}q_{I,l} &=0 & &\text{for $i,j,k,l$ distinct} \, .
\end{align}
Defining the vectors
\begin{align}
\vec{Q}_{i} &\equiv (q_{1,i}^2,q_{2,i}^2,\cdots,q_{X,i}^2) \\
\vec{A} &\equiv {1 \ov \sqrt{X}}(1,1,\cdots,1) \,,
\end{align}
and
\be
\vec{Q}_{ij} \equiv (q_{1,i}q_{1,j},q_{2,i}q_{2,j},\cdots,q_{X,i}q_{X,j}) \,
\ee
for $i< j$,
the equations obtained from \eq{e} and \eq{qnorm} can be re-written as,
\begin{align}
\vec{A} \cdot \vec{Q}_{i} &= \sqrt{108 \ov X} \\
\vec{Q}_{i}^2 &= 1 \\
\vec{Q}_{i} \cdot \vec{Q}_{j} =  \vec{Q}_{ij}^2 &= {1 \ov 3} & &\text{for $i,j$ distinct} \\
\vec{A} \cdot \vec{Q}_{ij} = \vec{q}_i \cdot \vec{q}_j &=0 & &\text{for $i,j$ distinct} \label{orth1} \\
\vec{Q}_{i} \cdot \vec{Q}_{ij} &=0 & &\text{for $i,j$ distinct} \label{orth2} \\
\vec{Q}_{i} \cdot \vec{Q}_{jk} = \vec{Q}_{ij} \cdot \vec{Q}_{ik} &=0 & &\text{for $i,j,k$ distinct} \label{orth3} \\
\vec{Q}_{ij} \cdot \vec{Q}_{kl} &=0 & &\text{for $i,j,k,l$ distinct} \label{orth4}
\end{align}
It is easy to see that $\vec{Q}_{i}$ and $\vec{Q}_{ij}$ are all non-zero,
since all $\vec{q}_i \neq \vec{0}$.
Also $\vec{Q}_{i}$ and $\vec{Q}_{ij}$ are $X$-dimensional
vectors by definition.

Using the given inner products we can show that
\begin{align}
|\vec{Q}_1 + \cdots + \vec{Q}_\n|^2
&=\n + {1 \ov 3} {\n (\n-1)}
={1 \ov 3} \n (\n+2) \,.
\end{align}
Since $\vec{A}$ is a unit vector by definition,
\be
\n \sqrt{108 \ov X} = \vec{A} \cdot (\sum_i \vec{Q}_{i} )
\leq |\sum_i \vec{Q}_{i}| = \sqrt{{1 \ov 3} \n (\n+2)} \,.
\ee
Hence
\be
{324\n \ov \n+2} \leq X \leq \n+273 \,, \label{mainapb1}
\ee
as promised.
This equation implies that
\be
\n \leq 17 \quad \text{or} \quad \n \geq 32 \,. \label{boundapb1}
\ee
An additional constraint is needed to obtain an
upper bound on $\n$.

The additional constraint can be obtained by utilizing the
full set of vectors $\vec{Q}_{ij}$ and $\vec{Q}_i$ we have defined.
Note that by \eq{orth3} and \eq{orth4}, $\vec{Q}_{ij}$ are mutually orthogonal.
They are also orthogonal to $\vec{Q}_i$ and $\vec{A}$
as can be seen in \eq{orth1}, \eq{orth2} and \eq{orth3}.

Also, all $\vec{Q}_i$ must be linearly independent.
This is because if we assume
\be
k_1 \vec{Q}_1 + \cdots + k_\n \vec{Q}_\n=0 \,,
\ee
then for non-zero $k_i$
\be
k_1^2 + \cdots k_\n^2 + {2 \ov 3} (k_1 k_2 + \cdots + k_{\n-1} k_\n)=0 \,.
\ee
But the l.h.s. can be rewritten as
\be
{2 \ov 3}(k_1^2 + \cdots k_\n^2) + {1 \ov 3} (k_1 + \cdots k_\n)^2 =0
\ee
and hence the equality cannot hold for non-zero $k_i$.

Therefore, we find that $\vec{Q}_i$ together with $\vec{Q}_{ij}$
form a set of linearly independent vectors.
This means that we must have $\n(\n+1)/2$ linearly independent vectors in
$X \leq \n+273$ dimensional space.
Hence,
\be
{\n(\n+1) \ov 2} \leq X \leq \n+273.
\ee
From this we obtain the bound $\n \leq 24$.

Put together with the bound \eq{boundapb1} we obtain
\be
\n \leq 17 \,,
\ee
as desired. $\Box$

\section{Proof of Minimal Charge Condition for $SU(13)\times U(1)$ Models}
\label{ap:prime}

In this section we prove that when
\begin{align}
r&=84n+43 = 2 \times 3 \times 7 \times (2n +1)+1\\
s&=182n+92 = 7 \times 13 \times (2n +1)+1
\end{align}
for integer $n$,
then the integers $a$ and $(-3a-2f/3)$ for
\begin{align}
a&=13r^2-234rs-51s^2 \\
f&=24(13r^2+3s^2)
\end{align}
are mutually prime.
Let us define
\be
g \equiv gcd(a,-3a-{2 \ov 3}f) = gcd (a,{2 \ov 3} f)\,.
\ee
Our goal is to show that $g=1$.

We first acknowledge that $r$ and $s$ are mutually prime.
This is because
\be
gcd(r,s)|(13 r -6s)
\ee
and
\be
13r-6s=7\,.
\ee
It is clear, however, that $7$ and $r$ are mutually prime.
Therefore, $gcd(r,s)$ must be 1, and hence $r$ and $s$ must be
mutually prime.
Meanwhile,
$a$ is odd since $r$ is even and $s$ is odd.
Therefore $g$ must also be odd, i.e. $2 \nmid g$.
Also, $g$ is not divisible by 3.
We can show $3 \nmid g$ by noting that
$g \mid (13r^2+3s^2)$ and that
\be
13r^2+3s^2 \equiv 1 \modular{3} \,.
\ee

Let us show that $g=1$. By definition
\begin{align}
\begin{split}
g &=  gcd (a,{2f \ov 3} ) = gcd(13r^2-234rs-51s^2, 16(13r^2+3s^2)) \\
&= gcd(13r^2-234rs-51s^2, 13r^2+3s^2) \,,
\end{split}
\end{align}
where we have used the fact that $2 \nmid g$.
Using standard properties of the greatest common divisor,
we further find that
\begin{align}
\begin{split}
g &= gcd(13r^2-234rs-51s^2, 13r^2+3s^2) \\
&= gcd(-234rs-54s^2,13r^2+3s^2) \\
&= gcd(-234r-54s,13r^2+3s^2)
= gcd(-18(13r+3s),13r^2+3s^2) \\
&= gcd(13r+3s,13r^2+3s^2)
\end{split}
\end{align}
In the penultimate line we have used the fact that
\be
\label{ess}
gcd(s,13r^2+3s^2) =gcd(s,13r^2) =1
\ee
since $s \equiv 1 \modular{13}$
and $s$ and $r$ are mutually prime.
In the last line we have used $2 \nmid g$
and $3 \nmid g$.

Therefore, $g$ must be a divisor of
\be
-(13r+3s)(13-3s)+13 (13r^2+3s^2)=48 s^2 \,.
\ee
We have seen in \eq{ess} that $s$ is mutually prime
with $13r^2+3s^2$. Therefore, $g$ is mutually prime with $s$
and hence
\be
g | 2^3 \times 3 \,.
\ee
We, however, know that $2 \nmid g$ and $3 \nmid g$.
This proves that $g=1$, as desired. $\Box$


\end{document}